\documentclass[useAMS,usenatbib,twocolumns]{mn2e}
\usepackage{graphicx}
\usepackage{multirow}
\usepackage{lipsum}
\usepackage{blindtext}
\usepackage{amsmath}
\usepackage{amssymb}
\usepackage{xspace}
\usepackage{pdflscape}
\usepackage{afterpage}
\usepackage{gensymb}
\usepackage{booktabs}
\usepackage{bold-extra}
\usepackage[usenames, dvipsnames]{color}
\usepackage{adjustbox}
\newcommand{\SAG}{{\sc sag}\xspace}
\newcommand{\SAGE}{{\sc sage}\xspace}
\newcommand{\GAL}{{\sc galacticus}\xspace}
\newcommand{\oiitit}{[O\scshape{ii}]\xspace}
\newcommand{\MD}{\textsc{MultiDark-Galaxies }}

\newcommand{\FOII}{F${\left[\mathrm{O\,\textrm{\textsc{ii}}}\right]}$\xspace}

\newcommand{\oem}{$\left[\ion{O}{{\scshape ii}}\right]$~emitters\xspace}
\newcommand{\OII}{$\left[\ion{O}{{\scshape ii}}\right]$\xspace}
\newcommand{\OIII}{$\left[\ion{O}{{\scshape iii}}\right]$\xspace}

\newcommand{\LOII}{$L{\left[\mathrm{O\,\textrm{\textsc{\scshape{ii}}}}\right]}$\xspace}
\newcommand{\OIIlamb}{$\left[\mathrm{O\,\textrm{\textsc{\scshape{ii}}}}\right]\lambda 3727$\xspace}
\newcommand{\OIIIlamb}{$\left[\mathrm{O\,\textrm{\textsc{\scshape{iii}}}}\right]\lambda 5007$\xspace}
\newcommand{\logaLOII}{log$_{10}(L{\left[\mathrm{O\,\textrm{\textsc{\scshape{ii}}}}\right]/\rm{erg\,s^{-1}}}$)\xspace}
\newcommand{\GE}{{\sc get\_\,emlines}\xspace}

\newcommand{\Mhalo}{$M_{\rm halo}\,$} 

\newcommand{\Mstar}{$M_{\star}\,$} 
\newcommand{\magu}{$M_u\,$} 
\newcommand{\magg}{$M_g\,$}

\title[\oem in MultiDark-Galaxies and DEEP2]{\vspace{-0.3cm}\hspace{0.5cm}[O{\Large{II}}]\xspace emitters in MULTIDARK-GALAXIES and DEEP2}
\author[Favole et al. 2020]
{\parbox[t]{\textwidth}{\vspace{-0.8cm}G. Favole,$^{1}$\thanks{E-mail: ginevra.favole@port.ac.uk} V. Gonzalez-Perez,$^{2,3,1}$\thanks{E-mail: violetagp@protonmail.com} D. Stoppacher,$^{4,5}$ \'A. Orsi,$^{6}$ J. Comparat,$^{7}$
S. A. Cora,$^{8,9}$ C. A. Vega-Mart\'inez,$^{10, 11}$ A. R. H. Stevens,$^{12}$ C. Maraston,$^{1}$ D. Croton,$^{13}$ A. Knebe,$^{5,12,14}$ A. J. Benson,$^{15}$ A. D. Montero-Dorta,$^{16}$ N. Padilla,$^{17,18}$ F. Prada,$^{19}$ D. Thomas$^{1}$}
\vspace*{15pt}\\
$^1$Institute of Cosmology and Gravitation, Portsmouth University, Burnaby Road, Portsmouth PO13FX, UK\\
$^2$Astrophysics Research Institute, Liverpool John Moores University, 146 Brownlow Hill, Liverpool L3 5RF, UK.\\
$^3$Energy Lancaster, Lancaster University, LA1 4YB, UK\\
$^4$Instituto de F\'{i}sica Te\'{o}rica (IFT) UAM/CSIC, Universidad Aut\'{o}noma de Madrid, Cantoblanco, E-28049 Madrid, Spain\\
$^5$Departamento de F\'isica Te\'orica, M\'odulo 15, Facultad de Ciencias, Universidad Aut\'onoma de Madrid, E-28049 Madrid, Spain\\
$^6$Centro de Estudios de F\'isica del Cosmos de Arag\'on, Plaza de San Juan 1, Teruel E-44001, Spain\\
$^7$Max-Planck-Institut f\"{u}r extraterrestrische Physik (MPE), Giessenbachstrasse 1, D-85748 Garching bei M\"{u}nchen, Germany\\
$^8$Instituto de Astrof\'isica de La Plata (CCT La Plata, CONICET, UNLP), Paseo del Bosque s/n, B1900FWA, La Plata, Argentina\\
$^{9}$Facultad de Ciencias Astron\'omicas y Geof\'isicas, UNLP, Paseo del Bosque s/n, B1900FWA, La Plata, Argentina\\
$^{10}$Instituto de Investigaci\'on Multidisciplinar en Ciencia y Tecnolog\'ia, Universidad de La Serena, Ra\'ul Bitr\'an 1305, La Serena, Chile\\
$^{11}$Departamento de Astronom\'ia, Universidad de La Serena, Av. Juan Cisternas 1200 Norte, La Serena, Chile\\
$^{12}$International Centre for Radio Astronomy Research, The University of Western Australia, Crawley, WA6009, Australia\\
$^{13}$Centre for Astrophysics \& Supercomputing, Swinburne University of Technology, P.O. Box 218, Hawthorn, Victoria 3122, Australia\\
$^{14}$Centro de Investigaci\'on Avanzada en F\'isica Fundamental (CIAFF), Facultad de Ciencias, UAM, 28049 Madrid, Spain\\
$^{15}$Carnegie Observatories, 813 Santa Barbara Street, Pasadena, CA 91101, USA\\
$^{16}$Instituto de F\'isica, Universidade de S\~ao Paulo, S\~ao Paulo, SP, Brazil\\
$^{17}$Instituto de Astrof\'isica, Pontificia Universidad Cat\'olica, Av. Vicu\~na Mackenna 4860, Santiago, Chile\\
$^{18}$ Centro de Astro-Ingenier\'ia, Pontificia Universidad Cat\'olica de Chile, Av. Vicu\~na Mackenna 4860, Santiago, Chile\\
$^{19}$Instituto de Astrof\'isica de Andaluc\'ia (IAA)/ CSIC, Granada, E-18008, Spain\\
\vspace{-1.2cm}}
\date{}
\begin{document}
\pagerange{\pageref{firstpage}--\pageref{lastpage}} \pubyear{2020}
\maketitle
\begin{abstract}
 We use three semi-analytic models (SAMs) of galaxy formation and evolution run on the same 1$h^{-1}$Gpc MultiDark Planck2 cosmological simulation to investigate the properties of \OII emission line galaxies at redshift $z\sim1$. We compare model predictions with different observational data sets, including DEEP2--{\sc Firefly} galaxies with absolute magnitudes. We estimate the \OII luminosity (\LOII) of our model galaxies using the public code \GE, which ideally assumes as input the instantaneous star formation rates (SFRs). This property is only available in one of the SAMs under consideration, while the others provide average SFRs, as most models do. We study the feasibility of inferring galaxies' \LOII  \,from average SFRs in post-processing. We find that the result is accurate for model galaxies with dust attenuated \LOII$\lesssim10^{42.2}$erg\,s$^{-1}$ ($<5\%$ discrepancy). 
 The galaxy properties that correlate the most with the model \LOII are the SFR and the observed-frame $u$ and $g$ broad-band magnitudes. Such correlations have r-values above 0.64 and a dispersion that varies with \LOII. We fit these correlations with simple linear relations and use them as proxies for \LOII, together with an observational conversion that depends on SFR and metallicity.
These proxies result in \OII luminosity functions and halo occupation distributions with shapes that vary depending on both the model and the method used to derive \LOII. The amplitude of the clustering of model galaxies with \LOII$>10^{40.4}$erg\,s$^{-1}$ remains overall unchanged on scales above 1$\,h^{-1}$Mpc, independently of the \LOII computation. 
\end{abstract}


\section{Introduction}
\label{sec:intro}
In the era of precision cosmology, surveys are starting to rely on star-forming galaxies to go further into early cosmic times, when dark energy is just starting to dominate the energy-matter budget of the Universe. Star-forming galaxies with strong nebular emission lines (ELGs) are among the preferred targets of the new generation of spectroscopic surveys as SDSS-IV/eBOSS \citep[][]{2016AJ....151...44D}, DESI \citep[][]{2015AAS...22533607S}, 4MOST\footnote{\url{https://www.4most.eu/cms/}}, WFIRST\footnote{\url{https://wfirst.gsfc.nasa.gov/}}, Subaru-PFS\footnote{\url{https://pfs.ipmu.jp/}} and Euclid\footnote{\url{http://sci.esa.int/euclid/}} \citep[][]{2011arXiv1110.3193L, 2015arXiv150502165S}, and will be used to trace the baryon acoustic oscillation \citep[BAO;][]{2005ApJ...633..560E} scale and the growth of structure by measuring redshift-space distortions in the observed clustering \citep[][]{2015ApJS..219...12A, 2018A&A...610A..59M, 2018orsi}. Star-forming galaxies will also be fundamental to inform halo occupation distribution \citep[HOD;][]{2002PhR...372....1C, Berlind2002, Kravtsov2004, Zheng2007} and (sub)halo abundance matching \citep[SHAM;][]{Conroy2006, Behroozi2010, Trujillo2011, 2013MNRAS.432..743N} models to generate fast mock galaxy catalogues useful for cosmological tests.

At $z\sim 1$ and for optical detectors, the samples of star-forming ELGs are dominated by \OII emitters. Therefore, measuring and modelling the relations between redshift and the physical properties of these galaxies -- such as \OII luminosity with star formation rate (SFR) --  is crucial for capitalising on the science that can be addressed from \OII data.  In this work, we aim to do exactly this, ultimately allowing us to build robust galaxy clustering predictions for near-future \OII data sets dominated by star-forming galaxies.

Modelling emission lines requires, at least, a certain knowledge of both the gas and the star formation history of a given galaxy. \OII emission is particularly difficult to predict, as it critically depends on local properties, such as dust attenuation, the structure of the H\,{\sc ii} regions and their ionisation fields. For this reason, \OII traces star formation and metallicity in a non-trivial way \citep[e.g.,][]{kewley04,mackay16}. Previous works on \oem have shown that semi-analytic models of galaxy formation (SAMs) are ideal laboratories for studying the physical properties of these galaxies, since they can reproduce the observed \OII luminosity function (LF) at $z \sim 1$ \citep[][]{orsi14,comparat2015,Comparat2016LFs,Hirschmann17}. \citet[][]{violeta2018} explored how \OII emitters are distributed in the dark matter haloes. They found typical host halo masses in agreement with the results from  \citet[][]{favole16}, which were based on a modified SHAM technique combining observational data with the MultiDark Planck dark matter N-body simulation \citep[MDPL;][]{klypin2016}.

For this project, we use the \MD mock products, which are publicly available at \url{https://www.cosmosim.org}. These catalogues were produced using 3 different SAMs to populate the snapshots of the MultiDark2 \citep[MDPL2;][]{klypin2016} dark matter cosmological simulation, over the redshift range $0<z<8$ \citep[][]{knebe2018}. MDPL2 is one of the largest dark matter simulation boxes with a volume of 1$\,h^{-3}$Gpc$^3$ and 3048$^3$ particles with mass resolution of $1.51\times10^9\,h^{-1}$M$\odot$. The models used in the production of these catalogues were: \SAG
\citep[][]{2015MNRAS.446.3820G, 2015MNRAS.446.2291M, 2018MNRAS.479....2C}, \SAGE \citep[][]{2016ApJS..222...22C} and \GAL \citep[][]{2012NewA...17..175B}. 

In this work, we explore the limitations of estimating the \OII luminosity in post-processing using different approaches, assessing how this quantity correlates with other galactic properties within the studied SAMs. The results from model galaxies are compared with observations from DEEP2 \citep[][]{newman2013}. The DEEP2 spectra have been fitted using \textsc{firefly} \citep[][]{firefly2017MNRAS,Comparat2017} to extract physical properties for these galaxies. Our final DEEP2 data set and SAM galaxy catalogues including emission line properties are made publicly available (see Sec.\,\ref{sec:dataavail}). 
In this analysis, we assume a \citet[][]{Planck15} cosmology with $\Omega_{\rm{m}}=0.6929$, $\Omega_{\Lambda}=0.3071$, $h=0.6777$.

The paper is organised as follows: in Section\,\ref{sec:data}, we describe the semi-analytic models considered in our study, the DEEP2 observational data set and the \textsc{firefly} code for spectral fitting. We compare the model SFR and stellar mass functions with current observations. In Section\,\ref{sec:O2data}, we describe how we calculate the \OII emission line luminosity in the SAMs using the publicly available code \GE by \citet[]{orsi14} with instantaneous SFR and cold gas metallicity as inputs. We analyse the impact of using average rather instantaneous SFR in this calculation to be used in those SAMs that do not provide instantaneous quantities. We compare the derived \OII luminosity functions with current observations.
In Section\,\ref{sec:proxies}, we explore the correlations between \LOII and several galactic properties to establish model proxies for the \OII luminosity. We provide scaling relations among these quantities that can be used in models without an emission line estimate. We further test these proxies by checking the consistency of the evolution of their \OII luminosity functions and clustering signal with observations and direct predictions from SAMs. Section\,\ref{sec:discussion} summarises our findings. 

\section{Data}
\label{sec:data}
\subsection{Semi-analytic models}
\label{sec:models}
Semi-analytic models of galaxy formation \citep[][]{1991ApJ...379...52W, kauffmann93} encapsulate the key mechanisms that
contribute to form galaxies in a set of coupled differential equations, allowing one to populate the dark matter haloes in cosmological $N$-body simulations with relative haste \citep[see e.g.,][]{baugh06,benson10,somerville15}. In the last two decades, a huge effort has been made to improve these models and account for the physical processes that shape galaxy formation and evolution, such as gas cooling \citep[e.g.,][]{delucia10,monaco14,hou17}, gas accretion \citep[e.g.,][]{guo11,henriques13,hirschmann16}, star formation \citep[e.g.,][]{lagos11}, stellar winds \citep[e.g.,][]{lagos13}, stellar evolution \citep[e.g.,][]{tonini09,henriques11,gp14}, AGN feedback \citep[e.g.,][]{bower06,croton2006}
or environmental processes \citep[e.g.,][]{2006MNRAS.372.1161W, font08, 2017MNRAS.471..447S, 2018MNRAS.479....2C}. 
Typically, SAMs do not attempt to resolve the scales on which these key astrophysical processes take place, but rather they describe their effects globally. Inevitably, this leads to free parameters in the models that require calibration; in essence, these compensate for the lack of understanding of certain processes and also for not resolving the relevant small scales.

In this study, we use the results from three semi-analytic models of galaxy formation: SAG \citep[][]{2006MNRAS.368.1540C, 2015MNRAS.446.3820G, 2015MNRAS.446.2291M, 2018MNRAS.479....2C}, SAGE \citep[][]{croton2006, 2016ApJS..222...22C} and Galacticus \citep[][]{2012NewA...17..175B}. The three SAMs considered have been run on the same MultiDark2 1$\,h^{-1}$Gpc cosmological simulation with Planck cosmology \citep[][]{klypin2016} to produce mock galaxy catalogues\footnote{publicly available at \url{https://www.cosmosim.org} and \url{http://skiesanduniverses.org/}}. 

The complete description of the first data release of the \MD products including \SAG, \SAGE and \GAL mock catalogues can be find in \citet{knebe2018}. All these catalogues lack \OII luminosity estimates. A version of Galacticus does have an emission line calculation \citep{Merson2018}, but it has not been applied to the MultiDark models. 

\subsubsection{\SAG}
\label{sec:SAG}
We consider a modified version of the Semi-Analytical Galaxies \citep[SAG;][]{2006MNRAS.368.1540C, 2008MNRAS.388..587L, 2015MNRAS.446.3820G, 2015MNRAS.446.2291M, 2018MNRAS.481..954C, 2018MNRAS.479....2C} code, which involves a detailed chemical model and implements an improved treatment of environmental effects (ram-pressure of both hot and cold gas phases and tidal stripping of gaseous and stellar components). It also includes the modelling of the strong galaxy emission lines in the optical and far-infrared range as described in \citet[][]{orsi14}. The free parameters of the model have been tuned by applying the Particle Swarm Optimisation technique \citep[PSO;][]{2015ApJ...801..139R} and using as constraints the stellar mass function at $z=0$ and 2 (data compilations from \citeauthor{2015MNRAS.451.2663H} \citeyear{2015MNRAS.451.2663H}), the SFR function at $z=0.14$ \citep[]{Gruppioni2015}, the fraction of mass in cold gas as a function of stellar mass \citep[]{2014A&A...564A..66B}, and the black hole--bulge mass relation \citep[]{2013ApJ...764..184M, 2013ARA&A..51..511K}.

\subsubsection{\SAGE}
\label{sec:SAGE}
The Semi-Analytic Galaxy Evolution\footnote{\url{http://www.asvo.org.au/}} code \citep[SAGE;][]{croton2006, 2016ApJS..222...22C} is a modular and customisable SAM. The updated physics includes gas accretion, ejection due to feedback, a new gas cooling--radio mode
AGN heating cycle, AGN feedback in the quasar mode, galaxy mergers, disruption, and the build-up of intra-cluster stars. 

\SAGE was calibrated to reproduce several statistical features and scaling relations of galaxies at $z=0$, including the stellar mass function, tightly matching the observational uncertainty range \citep[][]{2008MNRAS.388..945B}, the black hole-bulge mass relation, the stellar mass-gas metallicity relation, and the Baryonic Tully-Fisher relation \citep[]{1977A&A....54..661T}.  

\subsubsection{\GAL}
\label{sec:GAL}
\GAL\footnote{\url{https://bitbucket.org/galacticusdev/galacticus/wiki/Home}}
\citep[][]{2012NewA...17..175B} has much in common with the previous two models, in terms
of modularity, the range of physical processes included and the type of quantities that it can predict. Although this model has not been re-tuned to this simulation, the original calibration was performed using analytically built merger trees assuming a WMAP7 cosmology~\citep{2012NewA...17..175B}. The original model reproduced reasonably well the observed stellar mass function at $z\sim0.07$ \citep[][]{2009MNRAS.398.2177L} and the HI mass function at $z\sim0$ \citep[][]{martin2011}. 
\begin{figure}
\begin{center}
\includegraphics[width=0.85\linewidth]{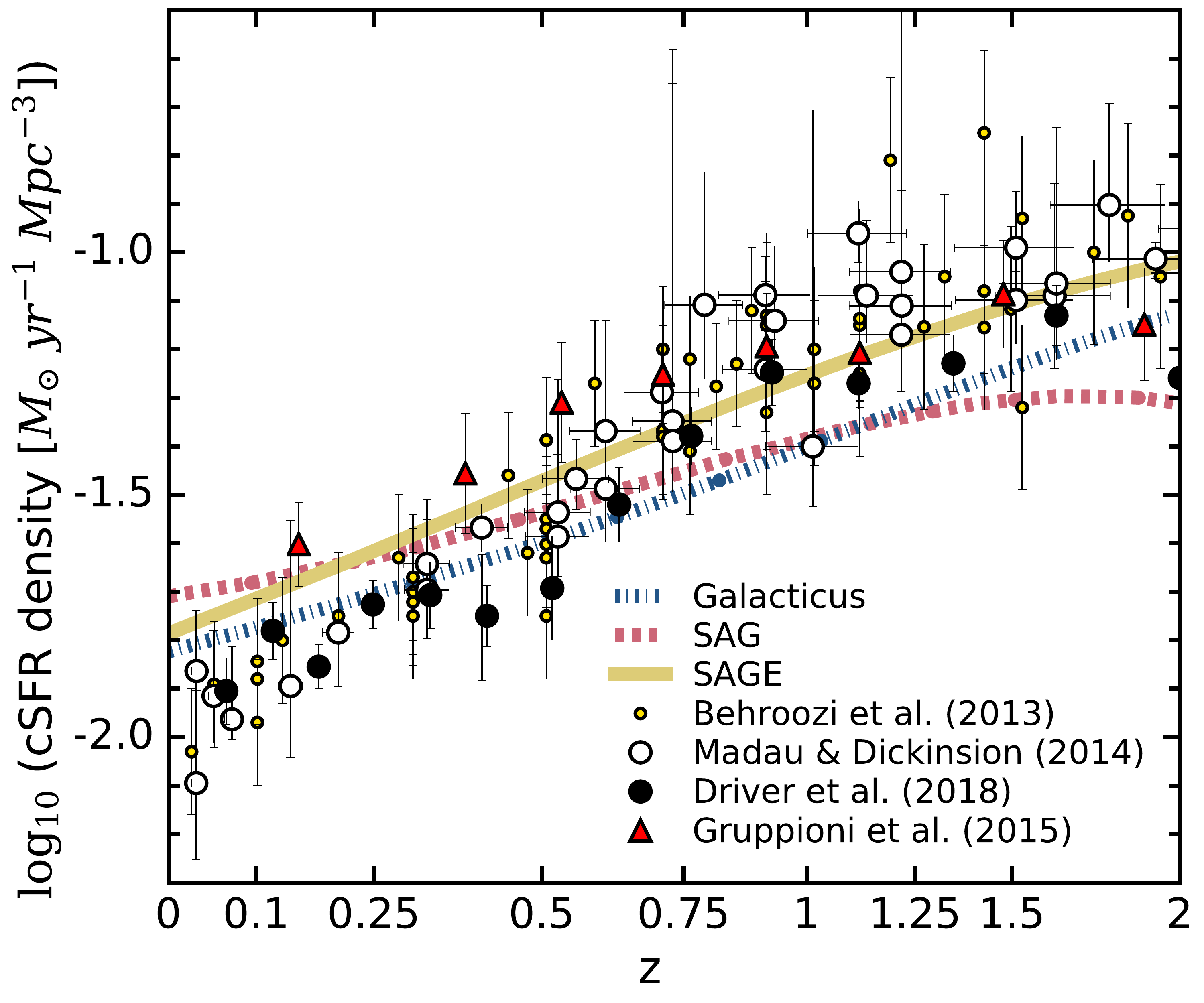}
\caption{Cosmic star formation rate density of \SAG, \SAGE and \GAL \MD as a function of redshift, compared to four independent compilations of data sets from \citet{Behroozi2013} (this was corrected to a \citet[]{Chabrier2014} IMF by the same authors), \citet{madau14}, \citet{driver18} and \citet{Gruppioni2015}. The error bars are the 1$\sigma$ dispersion around each point. We show this result only up to $z\sim2$, which is the maximum redshift of interest for our study. }
\label{fig:cSFR2}
\end{center}
\end{figure}

\subsubsection{Model comparison}
\label{sec:modelscompared}
For a full comparison between the \SAG, \SAGE and \GAL semi-analytic models adopted in this work, we refer the reader to \citet[][]{knebe2018}. 
The main differences between them are: i) the calibrations; ii) the treatment of mergers; iii) galaxies without a host halo, ``orphans", are not allowed in \SAGE, while they can happen, due to mass stripping, within \GAL and \SAG; and iv) the metal enrichment models, with \GAL and \SAGE assuming an instantaneous recycling approximation and \SAG implementing a more complete chemical model \citep{2006MNRAS.368.1540C, 2018MNRAS.481..954C}.

Here we also recall some results from \citet[][]{knebe2018} that are important for interpreting the outcomes of our analysis and a further study of global properties can be found in Appendix~\ref{sec:appendix2}. 
As we impose a minimum limit of \Mstar$>10^{8.87}\,$M$_{\odot}$ and SFR$\,>0\,\rm{yr^{-1}M_{\odot}}$ to the three SAMs of interest, some of our model results will be slightly different from those presented in \citet[][]{knebe2018}. The cuts above have been chosen taking into account the resolution limit of the MultiDark cosmological simulation \citep[see][]{knebe2018}. At $z\sim1$, the limit on SFR excludes about 4\% of the entire \SAG population, ~17\% of galaxies in \SAGE, and no galaxies from \GAL.

Fig.\,\ref{fig:cSFR2} shows the redshift evolution of the \MD cosmic star formation rate (SFR) density compared to a compilation of observations including estimates of the cosmic SFR from narrow-band (H$\alpha$), broad-band (UV-IR), and radio (1.4 GHz) surveys by \citet{Behroozi2013}, and more recent results by \citet{madau14}, \citet{Gruppioni2015} and \citet{driver18}. The observational data sets are consistent, despite being affected by different systematic errors. Fig.\,\ref{fig:cSFR2} only extends to $z\sim2$, as higher redshifts are not of interest for this study. All the SAMs agree with the observations within our redshift range of interest $0.6<z<1.2$. Beyond $z=2$, \SAG and \GAL model galaxies maintain a good agreement with the data out to $z\sim8.5$, while \SAGE overpredicts the SFR density at $z\gtrsim4$ \citep[see][]{knebe2018}. 
\begin{figure}
\centering
\includegraphics[width=\linewidth]{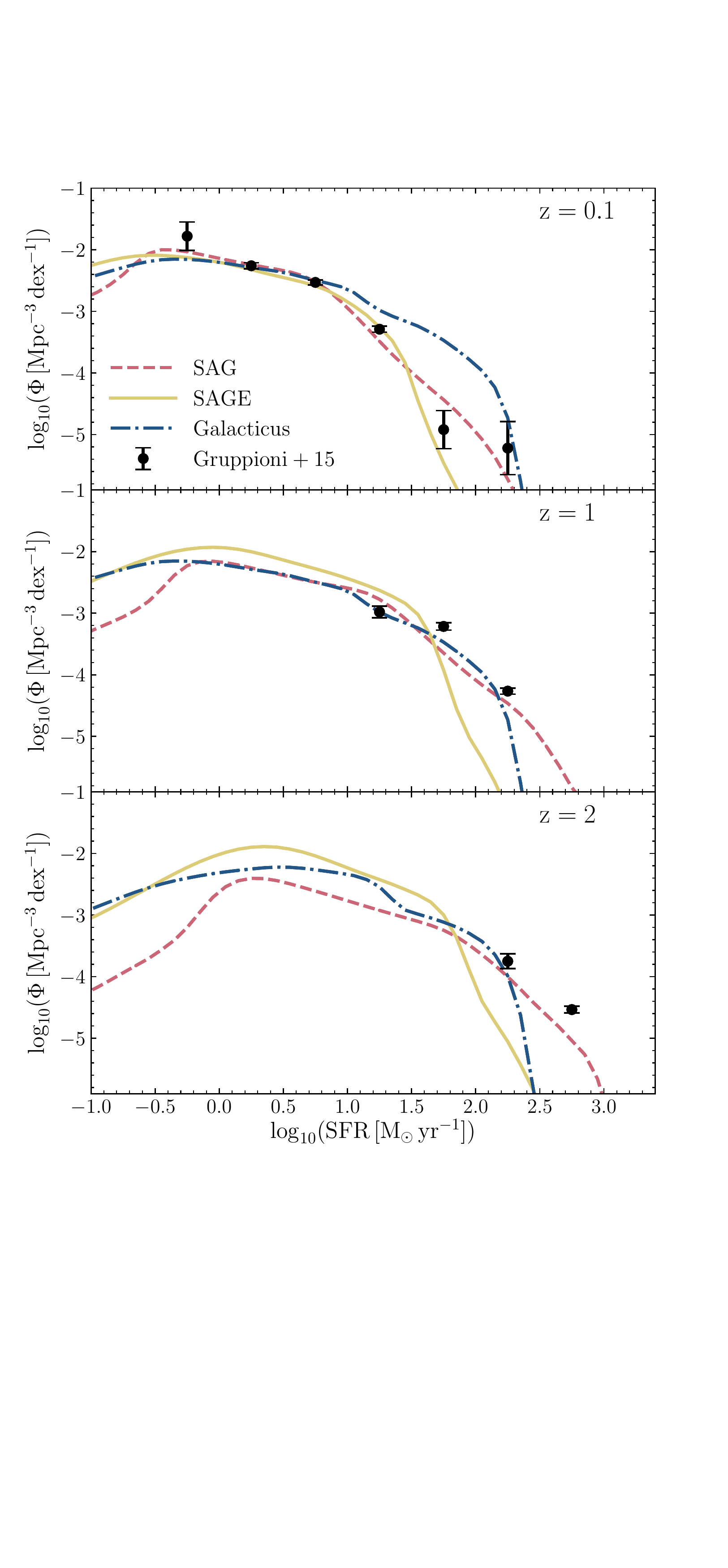}\vspace{-0.8cm}
\caption{\MD average SFR function evolution at $z\lesssim 2$ (lines) compared to the Herschel/PEP and HerMES observations \citep[][filled circles]{Gruppioni2015}.}
\label{fig:sfrf}
\end{figure}

In SAMs, galaxy properties are obtained by solving coupled differential equations in a certain number of steps in which the time interval between snapshots of the underlying DM simulation is divided. In this context, we define the ``instantaneous SFR'' as the star formation rate computed using the mass of stars formed over the last step before the output. The ``average SFR'' is instead the SFR obtained by considering the average contribution from all the steps. The \SAG model subdivides the time between snapshots in 25. This timescale typically corresponds to $\sim$10-25\,Myrs at $z\sim1$, which is the timescale physically relevant for the \OII emission. \SAGE and \GAL split time in 10 steps.

Fig.\,\ref{fig:sfrf} displays the average SFR functions of the \MD at different redshifts compared to the Herschel data from the PEP and HerMES surveys \citep[][]{Gruppioni2015}. We find good agreement for SAG model galaxies over the whole SFR and $z$ ranges considered. \GAL is consistent with the measurements at $\rm{SFR}\lesssim10^{2.5}\,$yr$^{-1}$M$_{\odot}$, while \SAGE agrees with the data up to $10^{2}\,$yr$^{-1}$M$_{\odot}$. At higher SFRs, \SAGE under-predicts the number of star-forming galaxies by $\sim2$\,dex.

\begin{figure}
\centering\vspace{-0.1cm}
\includegraphics[width=\linewidth]{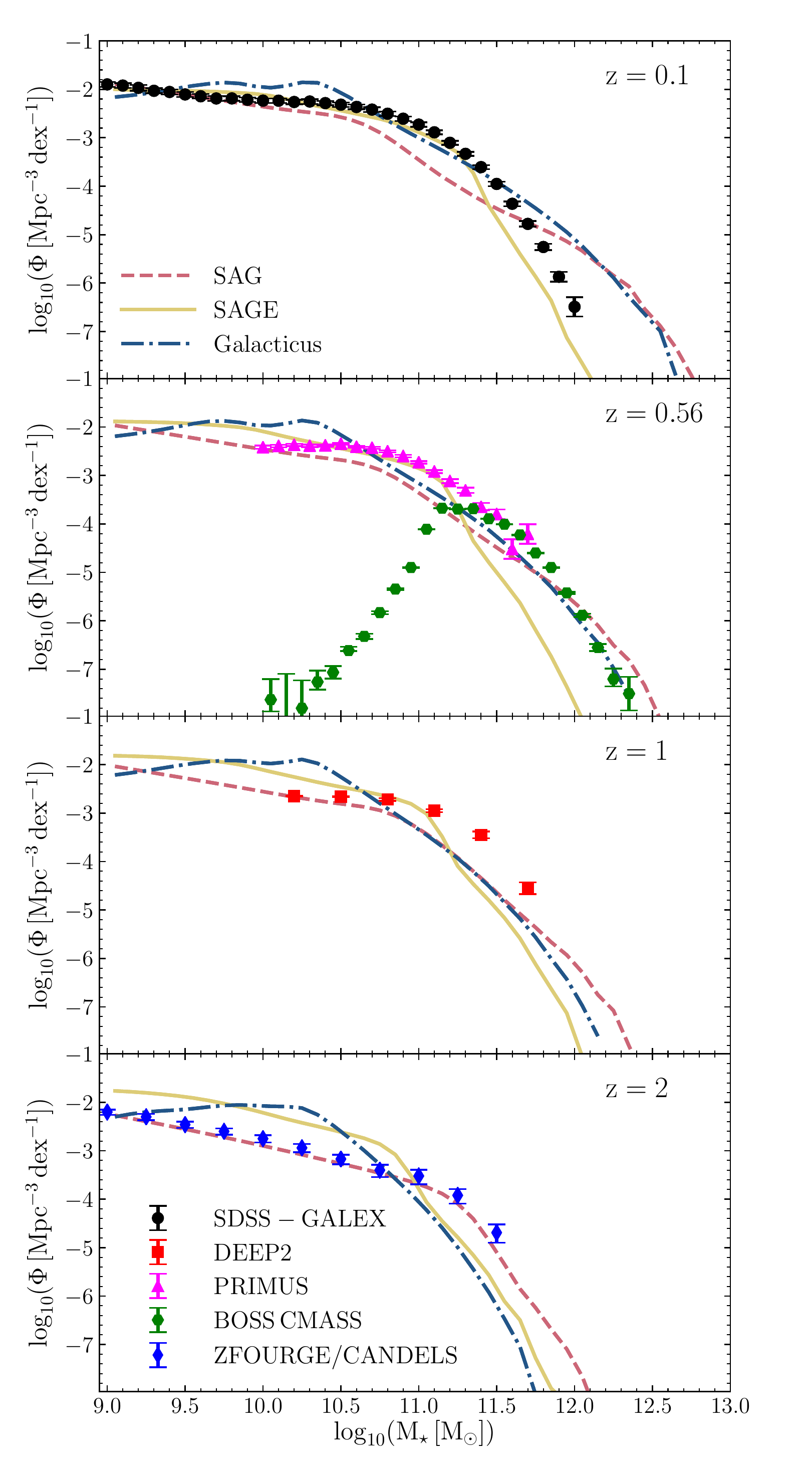}\vspace{-0.5cm}
\caption{Stellar mass function evolution of our model galaxies (lines colour-coded as in the legend) compared to the SDSS-GALEX $z=0.1$ \citep[][black points]{Moustakas2013} observations, the PRIMUS results at $0.50<z<0.65$ \citep[][magenta triangles]{Moustakas2013}, the BOSS CMASS measurements at $0.5<z<0.6$ not corrected from incompleteness \citep[][green hexagons]{maraston13}, the DEEP2-FF data at $0.9<z<1.1$ (red squares), and the ZFOURGE/CANDELS observations at $1.5<z<2.5$ \citep[][blue diamonds]{2014ApJ...783...85T}. Note that the BOSS data drop due to the selection of luminous, red, massive galaxies for this sample.}
\label{fig:massf}
\end{figure}

In Fig.\,\ref{fig:massf}, we show the evolution of the \MD stellar mass function compared to, from top to bottom, the SDSS-GALEX observations at $z=0.1$ \citep[][]{Moustakas2013}, the PRIMUS measurements at $0.50<z<0.65$ \citep[]{Moustakas2013}, the BOSS CMASS observations at $0.5<z<0.6$ \citep[][]{maraston13}, the DEEP2-FF data at $0.9<z<1.1$, and the ZFOURGE/CANDELS star-forming galaxies at $1.5<z<2.5$ \citep[][]{2014ApJ...783...85T}. The BOSS CMASS mass function drops in the low-mass end due to the incompleteness effect generated by the CMASS colour cuts specifically designed to select luminous, red, massive galaxies~\citep[][]{maraston13}. Note that the stellar mass functions shown in Fig.~\ref{fig:massf} are not the same as those from \citet[]{knebe2018} due to the SFR$>0$ cut we apply to the SAMs. The systematic errors on DEEP2 observations at $z\sim1$ are expected to differ from those of SDSS galaxies at lower redshifts.

It is not surprising that the agreement between \SAG and ZFOURGE/CANDELS data is especially good because this model was calibrated against these observations. \SAGE and \GAL over-predict the number of galaxies with $\rm{log(M_\star\,[M_\odot])\lesssim11}$, and this excess is enhanced at higher redshift (from $\sim0.1$\,dex at $z=0.1$ to $\sim0.4$\,dex at $z=2$). \SAGE under-estimates the number of galaxies more massive than $10^{11}$M$_\odot$ at all redshifts. We deem the \MD to be in sufficient agreement with observations in terms of their stellar-mass and SFR evolution such that we can draw meaningful predictions from the models that rely on these properties.

\subsection{DEEP2 galaxies}
\label{sec:deep2}
We are interested in exploring the relationship between \LOII and different galactic properties. For comparison, we use an observational data set, the DEEP2--\textsc{Firefly} (DEEP2-FF, hereafter)  galaxy sample, which allows us to test whether the model galaxies cover similar ranges of parameters, once the adequate selection functions are implemented.

The DEEP2 survey obtained spectra of about 50,000 galaxies brighter than $R\sim 24.1$, in four separate fields covering $\sim2.8$\,deg$^2$ \citep[][]{newman2013}. The redshift measurement for each object in the DEEP2 DR4 database was inspected by eye and assigned an integer quality code $-2<Q<4$ based on the determined accuracy of the redshift value.\footnote{\url{http://deep.ps.uci.edu/DR4/zquality}} For this work, we consider galaxies with $Q>2$, corresponding to secure redshifts, within the range $0.001<z_{\rm {best}}<1.7$.  

We adopt the DEEP2 flux-calibrated spectra generated by \citet[][]{Comparat2016LFs}.\footnote{\url{http://www.mpe.mpg.de/~comparat/DEEP2/}} We also use the extended photometric catalogues developed by \citet[][]{matthews2013},\footnote{\url{http://deep.ps.uci.edu/DR4/photo.extended}} which supplement the DEEP2 photometric catalogues with ($u,g,r,i,z$) photometry from the Sloan Digital Sky Survey (SDSS). By applying the cuts specified above and taking into account the cross-match between the mentioned catalogues, the spectra of 33\,838 galaxies from the original DEEP2 DR4 catalogue are used in this study. These spectra are fitted using stellar population models to extract quantities such as stellar masses, stellar metallicities, star formation rates, and ages. In particular, the DEEP2 SFR values are computed by fitting stellar population models to the spectral continuum, where the emission lines are masked for the fit. Thus, this constitutes an independent estimate from an \OII-based SFR.

The spectral fit is performed using the \textsc{firefly}\footnote{\url{https://github.com/FireflySpectra/Firefly_release},\url{http://www.icg.port.ac.uk/Firefly/}} code \citep[][]{firefly2017MNRAS, Comparat2017} in which no priors, other than the assumed model described immediately below,
are applied. \textsc{firefly} treats dust attenuation in a novel way, by rectifying the continuum before the fit; for full details see \citet[][]{firefly2017MNRAS} and \citet[][]{Comparat2017}. The \textsc{firefly} fit is performed for spectral templates with ages below 20 Gyr and metallicities
in the range $0.001<Z<3$. The maximum age found for the DEEP2-FF sample is 10.18 Gyr. It is noteworthy to remark that \textsc{firefly} does not interpolate between the ages of the templates used in the spectral fitting.
For this study, we adopt spectral templates from \citet[][]{Maraston2011}, assuming a \citet[][]{Chabrier2003} IMF, same as in the \textsc{MultiDark-Galaxies}, and the ELODIE stellar library. This latter covers the wavelength range $3900-6800$\,\AA \,with a 0.55\,\AA \,sampling at 5500\,\AA, i.e. at a resolution $R=10,000$ \citep[][]{Prugniel2007}.

The DEEP2 survey used the DEIMOS spectrograph at Keck, which covers approximately the wavelength range $6500-9300$\,\AA\, with a resolution $\sim\,$6000 \citep[][]{Faber2003}. The discrepancy in wavelength coverage results in a lack of fits at low redshifts for this survey.

The \textsc{firefly} fits to the DEEP2 spectra described above are available at \url{http://www.icg.port.ac.uk/Firefly/} (340 MB). Another fit to the DEEP2 spectra has been performed by \citet[][]{Comparat2017} assuming slightly different age and metallicity ranges, and using a previous version of \textsc{firefly} that did not take into account the presence of mass loss in the stellar population models. Here we refer to ``stellar mass" as the sum of the mass of living stars and the mass locked in stellar remnants (i.e., white dwarfs, neutron star and black holes).

\subsubsection{Broad-band absolute magnitudes}
The DEEP2-FF galaxy catalogue also provides SDSS $(u,g,r,i,z)$ apparent magnitudes. In order to compare these observations with the \MD absolute magnitudes, we have $(k+e)$ corrected them (where ``$e$'' stands for evolution). To this end, we have produced an evolving set of simple stellar populations
\citep[SSP;][]{Maraston2011} with ages, metallicities, and redshifts matching those used for the \textsc{Firefly} runs described above. In particular, we produce a table of possible evolutionary paths that provides the observed-frame properties of the given SSPs in the SDSS filters and allows us to determine the $k$-correction in those filters without any approximation. Hereafter, we will call it \textquoteleft\textquoteleft MS table''. This table calculates intrinsic magnitudes. The DEEP2 data have been corrected from interstellar dust attenuation by applying \citet[]{Calzetti2000} extinction law. 

These SDSS observed-frame properties are computed  by red-shifting the model SEDs to a fixed grid of redshifts from $z=3.5$ down to $z=0.$, with $\Delta z=0.1$, and applying cosmological dimming using the Flexible-k-and-evolutionary-correction algorithm (\textsc{FLAKE}, {\color{blue}{Maraston, in prep.}}). We interpolate between the redshifts when needed. Such a technique has been widely used in the literature \citep[e.g.,][]{maraston13, 2017MNRAS.466..228E} and can be generalised to any arbitrary set of filters. 

From each SSP model in the MS table above we extract the $(k+e)$ correction as:
\begin{equation}
(k+e)_{j}=M_j(z)-m_j=M_j(z)-M_j(z=0),
\label{eq:kecorr}
\end{equation}
where $M_j(z)$ are the galaxy SDSS $j=(u,g,r,i,z)$ absolute magnitudes at redshift $z$ and $m$ are the observed magnitudes, i.e. the absolute magnitudes at $z=0$.

The \textsc{Firefly} spectral fitting code finds the best fit to a galaxy by weighting different SSPs and adding them together. It turns out that the best \textsc{Firefly} fits to the DEEP2 galaxy sample have only two SSP components. Thus, the DEEP2-FF galaxy sample can be cross-matched with the components of the MS table, by using a linear combination of the two SSP components of each \textsc{Firefly} (FF) best fit:
\begin{equation}
{\rm{SSP}}^{\rm{MS}}=w_0\,{\rm{SSP}}_0^{\rm{FF}} + w_1\,{\rm{SSP}}_1^{\rm{FF}},
\end{equation}
with $w_0+w_1=1$. Then, each DEEP2-FF galaxy is assigned a $(k+e)$ correction that is the weighted, linear combination of the corrections from each SSP component:
\begin{equation}
(k+e)_{j}=(k+e)_{j}^0\,w_0+(k+e)_{j}^1\,w_1.
\end{equation}

\subsubsection{The DEEP2--{\sc FIREFLY} galaxy sample}
\label{sec:samplefinale}
For our analysis, we focus on DEEP2-FF galaxies within the redshift range $0.9<z<1.1$. We consider the sum of the \OII 3727\AA\,and 3729\AA\,line fluxes as the \OII doublet. Here we impose a flux limit of \FOII$\,\rm{>5\,\sigma_{F_{[OII]}}}$ (where $\sigma_{F_{[OII]}}$ is the flux error) to guarantee robust flux estimates, and a minimum stellar mass uncertainty of $\rm{[\log_{10}(M_{\star}^{1\sigma\,up})-\log_{10}(M_{\star}^{1\sigma\,low})]/2<0.4}$. In the previous expression, $\rm{M_{\star}^{1\sigma\,up,low}}$ represents the \textsc{Firefly} stellar mass within $\pm1\,\sigma$ from the mean value of the distribution. 

After applying the cuts described above, our final sample includes 4478 emitters with minimum \OII flux of $2.45\times10^{-19}$\,erg\,s$^{-1}$\,cm$^{-2}$, mean \LOII$\sim10^{41.6}$\,erg\,s$^{-1}$, \Mstar$\sim10^{10.3}\,$M$_{\odot}$, age$\,\sim10^{9.2}\,$yr, and mean cold gas metallicity $Z_{\rm{cold}}\sim0.72$. Fig.\,\ref{fig:SFRinstavg} shows the distribution of \LOII as a function of SFR. The observed sample only populates a narrow range of SFR, and this affects the comparison with the model galaxies, which have SFRs lower than the minimum value of the DEEP2-FF sample. Other properties from this data set can be seen in Fig.\,\ref{fig:proponly} and in Appendix\,\ref{sec:appendix2}.
We assume the dust attenuation of the nebular emission lines to be the same as for the continuum. Thus, we also correct the \LOII from interstellar dust attenuation by applying \citet[]{Calzetti2000} extinction law, as we have detailed above for the broad-band magnitudes.

For the analysis, we select both observed and models galaxies using a more conservative flux cut, \FOII$>5\times10^{-18}$\,erg\,s$^{-1}$\,cm$^{-2}$. This corresponds to \LOII$\sim10^{40.4}$\,erg\,s$^{-1}$ at $z=1$ in Planck cosmology \citep[][]{Planck15}, and roughly mimics the observational limitations \citep[see also][]{violeta2018}. This cut reduces the sparse, faint tail of the observed distribution (there are only 4 DEEP2 galaxies with flux lower than $5\times10^{-18}$\,erg\,s$^{-1}$\,cm$^{-2}$) and allows us to obtain much narrower SAM constraints. 

\begin{figure}
\centering\vspace{-0.8cm}
\includegraphics[width=\linewidth]{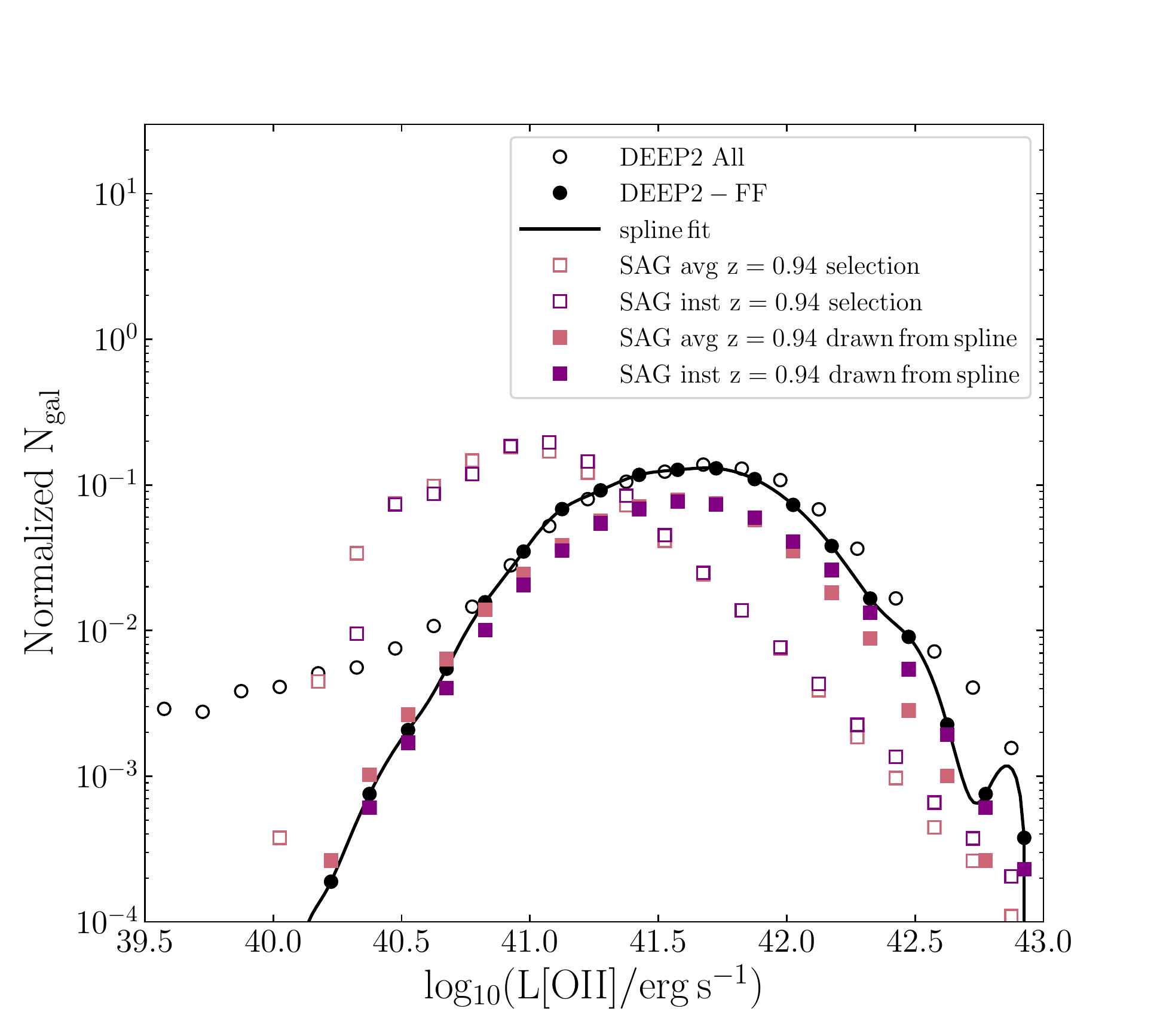}\vspace{-0.3cm}
\caption{Attenuated \LOII distribution of the original DEEP2 sample (black, empty dots) compared to the DEEP2-FF selection (black, filled dots), which we fit with a spline function.  The area under the curves is normalized to unity. We compare these results with the \SAG model galaxies selected with \Mstar$>10^{8.87}{\rm M}_{\odot}$ and SFR$>0\,\rm{yr^{-1}M_{\odot}}$ (empty squares; see Sec.\,\ref{sec:modelscompared}), and with the SAG galaxies randomly drawn from the DEEP2-FF spline distribution (filled squares). The \OII luminosity values in the model galaxies are calculated using the \GE code, inputing either the instantaneous (purple) or average (salmon) SFR. All the details about these quantities and the calculations are given in Sec.\,\ref{sec:O2sams}.}
\label{fig:spline}
\end{figure}

As shown in Fig.\,\ref{fig:spline}, most galaxies with \LOII$<10^{40.5}$\,erg\,s$^{-1}$ have been removed from the DEEP2-FF sample, compared to the original DEEP2 population. Despite this, the DEEP2-FF galaxy sample is statistically representative of the original DEEP2 population. In fact, the cumulative distribution functions of these two samples, approximated by splines, differ by less than 5\%, according to a Kolmogorov–Smirnov test.

Fig.\,\ref{fig:spline} shows the distribution of the dust-attenuated \LOII computed with \GE (see \S\ref{sec:O2sams}) from instantaneous and average SFRs, for SAG model galaxies at $z=0.94$ and with \Mstar$>10^{8.87}{\rm M}_{\odot}$ and SFR$>0\,\rm{yr^{-1}M_{\odot}}$ (see Sec.\,\ref{sec:modelscompared}). These model \LOII distributions are statistically different from the DEEP2-FF one. However, they have similar mean values: $\langle L{\left[\mathrm{O\,\textrm{\textsc{{\scshape ii}}}}\right]}\rangle\sim10^{41}{\rm erg\,s^{-1}}$, $\langle \rm{age}\rangle\sim 10^{9.2}$yr, $\langle M_{\star}\rangle\sim10^{9.47}{\rm M}_{\odot}$.

In order to draw a sample of model galaxies consistent with DEEP2-FF observations, we select \SAG galaxies with a \LOII distribution following the spline fit to the DEEP2-FF distribution, as shown in Fig.\,\ref{fig:spline}. We perform such a drawing for \SAG \OII luminosities computed both from instantaneous and average SFR. The \LOII of these new selections have mean values consistent with those from the DEEP2-FF sample. Meanwhile, the ages, $\langle \rm{age}\rangle\sim 10^{9}$yr, and the stellar masses, $\langle M_{\star}\rangle\sim10^{9.86}{\rm M}_{\odot}$, are lower than the observed ones.

In Appendix~\ref{sec:spline_results}, the DEEP2-FF sample is directly compared to the \SAG model galaxies selected following the DEEP2-FF \LOII distribution. These \SAG model subsets have brighter \LOII, M$_u$ and M$_g$ values, slightly lower ages, higher stellar masses and span higher SFR values compared to the SAG selection at \Mstar$>10^{8.87}{\rm M}_{\odot}$ and SFR$>0\,\rm{yr^{-1}M_{\odot}}$. 

The main focus of this paper is to test the validity of different approaches for modelling emission lines in large galaxy samples with volumes comparable to the observable Universe. In this context, the comparison to the DEEP2-FF sample is meant to be a rough guide to the expected location of observed galaxies in different parameter spaces.

\section{{\oiitit} emitters in the SAMs}
\label{sec:O2data}
The physics of \OII emission lines is difficult to model, as it depends on local processes, such as dust extinction, and the inner structure and the ionising fields of the H\,{\sc ii} nebula in which they are embedded. 
Different approaches have been used to model the \OII emission line: (i) assume a relation between \LOII and SFR and, possibly, metallicity as it happens in observations \citep[]{1998ARA&A..36..189K,kewley04, 2006ApJ...642..775M,jouvel2009,2012MNRAS.420.1926S,talia2015,valentino2017}; (ii) assume an average H\,{\sc ii} region for a range of metallicities \citep{violeta2018}; (iii) couple a photoionisation model with a galaxy evolution one \citep{2012MNRAS.419.3200H,orsi14}. We address method (i) in Section\,\ref{sec:proxies} and method (iii) here.

None of the \MD catalogues studied in this work provides direct \LOII estimates. Therefore, we couple the SAMs with the \GE model \citep[]{orsi14}, which encapsulates the results from the MAPPINGS-III photoionisation code \citep[][]{mappings1, mappings2}. Here, the ionisation parameter of gas in galaxies is directly related to their cold gas metallicity, obtaining a reasonable agreement with the observed H$\alpha$, \OIIlamb, \OIIIlamb luminosity functions, and the Baldwin, Phillips \& Terlevich \citep[BPT;][]{1981PASP...93....5B} diagram for local star-forming galaxies. 
Ideally, the \GE methodology requires as input the cold gas metallicity and the instantaneous SFR. This latter quantity, however, is not usually output by SAMs. The instantaneous SFR is preferred to a time-averaged equivalent, as the latter can include contributions from stellar populations older than those responsible for generating the nebular emission in star-forming galaxies.

\SAG is the only semi-analytic model providing both instantaneous and average SFR values, while \SAGE and \GAL only provide average SFRs.
In the next section, we describe in detail the \GE algorithm to be used in the \LOII calculation for a semi-analytic model. Because SAMs do not usually output the instantaneous SFR, which is needed as default input for the \GE code, we test the usage of the average SFR and how this affects different galactic properties.

\subsection{The \textsc{GET\_\,EMLINES} code}
\label{sec:O2sams}
We now describe step by step how we have implemented the \GE nebular emission code to obtain \OII luminosities for the \textsc{MultiDark-Galaxies}.
This methodology is  based on the photoionisation code MAPPINGS-III \citep[][]{mappings1, mappings2}, which relates the ionisation parameter of gas in galaxies, $q$, to their cold gas metallicity $Z_{\rm{cold}}$ as:
\begin{equation}
    q(Z)=q_0\left(\frac{Z_{\rm{cold}}}{Z_0}\right)^{-\gamma},
    \label{eq:ionpar}
\end{equation}
where $q_0$ is the ionisation parameter of a galaxy that has cold gas metallicity $Z_0$ and $\gamma$ is the exponent of the power law. Following \citet[]{orsi14}, from the pre-computed H\,{\sc ii} region model grid of \citet[]{2010ApJ...712L..26L}, we assume $q_0=2.8\times10^7\,\rm{cm\,s^{-1}}$, {\bf $Z_0=0.012$} and $\gamma=1.3$ for all the analysed galaxy models. This specific combination of values was presented in \citet[]{orsi14}, and it has ionization parameter values that bracket the range spanned by the MAPPINGS-III grid for the bulk of the galaxy population studied in that work.
The $q_0$ and $\gamma$ parameters above were found to produce model H$\alpha$, \OIIlamb (to indicate the doublet), \OIIIlamb luminosity functions and a model BTP \citep[][]{1981PASP...93....5B} diagram for local star-forming galaxies in good agreement with observations. 

The \GE code has been calibrated to reproduce a range of luminosity functions at different redshifts and local line ratios diagrams, and it has been tested against observations up to $z=5$~\citep{orsi14}. A different combination of $q_0$ and $\gamma$ changes the \LOII results in a complicated way. For instance, higher parameter values produce a lower number density of bright emitters, which translates into a substantial difference in the lower peak of the \LOII-SFR bimodality shown in Fig.\,\ref{fig:SFRinstavg}. Changing the $q_0$ and $\gamma$ parameters would require to recalibrate the \GE model, and this goes beyond the scope of this work. 

The cold gas metallicity is defined as the ratio between the cold gas mass in metals and the cold gas mass \citep[e.g.,][]{Yates2014}, considering both bulge and disc components, when available:
\begin{equation}
Z_{\rm{cold}}=\frac{M_{Z\rm{cold}}}{M_{\rm{cold}}}.
\label{eq:metallicity}
\end{equation}

Another fundamental quantity needed to derive the \OII line luminosity is the hydrogen ionising photon rate defined as:
\begin{equation}
    Q_{\rm{H^0}}=\int_0^{\lambda_0}\frac{\lambda L_\lambda}{hc}d\lambda,
    \label{eq:ionphotrate}
\end{equation}
where $L_\lambda$ is the galaxy composite SED in erg\,s$^{-1}$\,\AA$^{-1}$, $\lambda_0=912$\AA, $c$ is the speed of light and $h$ is the Planck constant. $Q_{\rm{H^0}}$ is a unit-less quantity calculated at each model snapshot just by solving the integral above. Assuming a \citet[]{2001MNRAS.322..231K} IMF, one can express the ionising photon rate as a function of the instantaneous star formation rate as \cite[]{2013seg..book.....F}:
\begin{equation}
Q_{\rm{H^0}} = \rm{log_{10}1.35 + log_{10}(SFR/M_\odot\,yr^{-1}) + 53.0.}
\label{eq:ionphotrate_shortcut}
\end{equation}

Combining Eq.\,\ref{eq:ionphotrate_shortcut} with the attenuation-corrected emission-line lists from \citet[]{2010ApJ...712L..26L}, normalised to the H$\alpha$ line flux, we compute the \OII luminosity as:
\begin{equation}
L(\lambda_j)=1.37\times10^{-12}Q_{\rm{H^0}}\frac{F(\lambda_j,q,Z_{\rm{cold}})}{F(H\alpha,q,Z_{\rm{cold}})},
\label{eq:lum}
\end{equation}
where $F(\lambda_j,q,Z_{\rm{cold}})$ is the MAPPINGS-III prediction of the desired emission line flux at wavelength $\lambda_j$ for a given set of ($q$, $Z_{\rm{cold}}$) and $F(H\alpha,q,Z_{\rm{cold}})$ is the H$\alpha$ normalisation flux.

The total luminosity of the \OII doublet is the sum of the luminosities of the two lines at $\lambda_j=3727,3729$\,\AA, both calculated using Eq.\,\ref{eq:lum}. 

The \OII luminosity in Eq.\,\ref{eq:lum} does not include any dust contribution. In order to account for dust attenuation, we implement the correction detailed in next Section using \citet[][]{Cardelli1989} extinction curve.


\subsection{Dust extinction}
\label{sec:dust}
In this study, the intrinsic \OII luminosity given in Eq.\,\ref{eq:lum}, $L(\lambda_j)$, is attenuated by interstellar dust as follows:
\begin{equation}
    L(\lambda_j)^{\rm{att}}=L(\lambda_j)10^{-0.4A_\lambda(\tau_\lambda^z,\theta)},
    \label{eq:attenuation}
\end{equation}
where $A_\lambda(\tau_\lambda^z,\theta)$ represents the attenuation coefficient defined as a function of the galaxy optical depth $\tau_\lambda^z$ and the dust scattering angle $\theta$. Explicitly we have \citep[]{1978ppim.book.....S, 1989agna.book.....O, 2003ARA&A..41..241D, 2019MNRAS.488..609I}:
\begin{equation}
A_\lambda(\tau_\lambda^z,\theta)=-2.5\log_{10}\frac{1-\exp(-a_\lambda\,\sec\theta)}{a_\lambda\,\sec\theta},
\label{eq:attcoeff}
\end{equation} 
where $a_\lambda=\sqrt{1-\omega_\lambda}\tau_\lambda^z$ and $\omega_\lambda$ is the dust albedo, i.e. the fraction of the extinction that is scattering. We assume $\cos\theta=0.60$ and $\omega_\lambda=0.80$, meaning that the scattering is not isotropic but more forward-oriented, and that 80\% of the extinction is scattering. These are the values that return the best agreement with DEEP2+VVDS observations in Fig.\,\ref{fig:LFz}.

The galaxy optical depth $\tau_\lambda^z$ that enters Eq.\,\ref{eq:attcoeff} is defined as \citep[][]{1999A&A...350..381D, 2003MNRAS.343...75H, 2007MNRAS.375....2D}:
\begin{equation}
\tau_{\lambda}^z=\left( \frac{A_{\lambda}}{A_V}\right)_{Z_{\odot}}\left( \frac{Z_{\rm{cold}}}{Z_{\odot}}\right)^s\left( \frac{\langle N_H\rangle}{2.1\times10^{21}\rm{atoms\,\,cm^{-2}}}\right),
\label{eq:tau}
\end{equation}
where the first two factors on the right-hand side represent the extinction curve. This depends on the cold gas metallicity $Z_{\rm{cold}}$ defined in Eq.\,\ref{eq:metallicity} according to power-law interpolations based on the solar neighbourhood, the Small and the Large Magellanic Clouds. The exponent $s=1.6$  \citep[][]{Guiderdoni1987} holds for the $\lambda>2000$\AA\, regime, where the \OII line is located. The $\left( A_{\lambda}/A_V\right)_{Z_{\odot}}$ term is the extinction curve for solar metallicity, which we take to be that of the Milky Way, and $\langle N_H\rangle$ the mean hydrogen column density. We adopt the values $Z_{\odot}=0.0134$ \citep[]{2009ARA&A..47..481A} for the solar metallicity.

Assuming the \citet[][]{Cardelli1989} extinction law in $0.3\,\mu\rm{m}\leq\lambda<0.9\,\mu\rm{m}$ (i.e., optical/NIR regime), one has:
\begin{equation}
\left(\frac{A_\lambda}{A_V}\right)=a(x)+b(x)/R_V, 
\label{eq:extcurve}
\end{equation} 
where $x\equiv\lambda^{-1}$, $R_V\equiv A_V/E(B-V)=3.1$ is the ratio of total to selective extinction for the diffuse interstellar medium in the Milky Way, and
\begin{equation}
\begin{aligned}
a(x)=&1+0.17699\,y-0.50447\,y^2-0.02427\,y^3+\\
      &0.72085\,y^4+0.01979\,y^5-0.77530\,y^6+0.32999\,y^7,\\
      b(x)=&1.41338\,y+2.28305\,y^2+1.07233\,y^3-5.38434\,y^4\\
      &-0.62251\,y^5+5.30260\,y^6-2.09002\,y^7,
\end{aligned}
\end{equation} 
with $y=(x-1.82)$.

The mean hydrogen column density is given by \citep[]{2003MNRAS.343...75H, 2007MNRAS.375....2D}:
\begin{equation}\label{eq:columndens}
\langle N_H\rangle=\frac{M_{\rm{cold}}^{\rm{disc}}}{1.4\,m_p\,\pi\,(a\,R_{1/2}^{\rm{disc}})^2}\,\,{\rm{atoms\,\,cm^{-2}}},
\end{equation}
where $M_{\rm{cold}}^{\rm{disc}}$ is the cold gas mass of the disc, $m_p=1.67\times10^{-27}\,{\rm{kg}}$ is the proton mass, $a=1.68$ is such that the column density represents the mass-weighted mean column density of the disc, and $R_{1/2}^{\rm{disc}}$ is the disc half-mass radius. 

Qualitatively for this dust attenuation model\footnote{Our implementation of the dust attenuation model is available at \url{https://github.com/gfavole/dust}}, galaxies with large amounts of cold gas, metal rich cold gas and/or small scale sizes, will be the most attenuated ones~\citep[see also][]{merson2016}.

\subsection{Instantaneous versus average SFR}
\label{sec:instavg}
The \GE code described in Section\,\ref{sec:O2sams} ideally requires as inputs the instantaneous SFR and cold gas metallicity of galaxies. The instantaneous SFR, which is defined on a smaller time-step compared to the average SFR (see Sec.\,\ref{sec:modelscompared}), traces very recent or ongoing episodes of star-formation, that are the relevant ones for nebular emission.

\begin{figure}
\begin{center}\vspace{-0.8cm}
\includegraphics[width=1.05\linewidth]{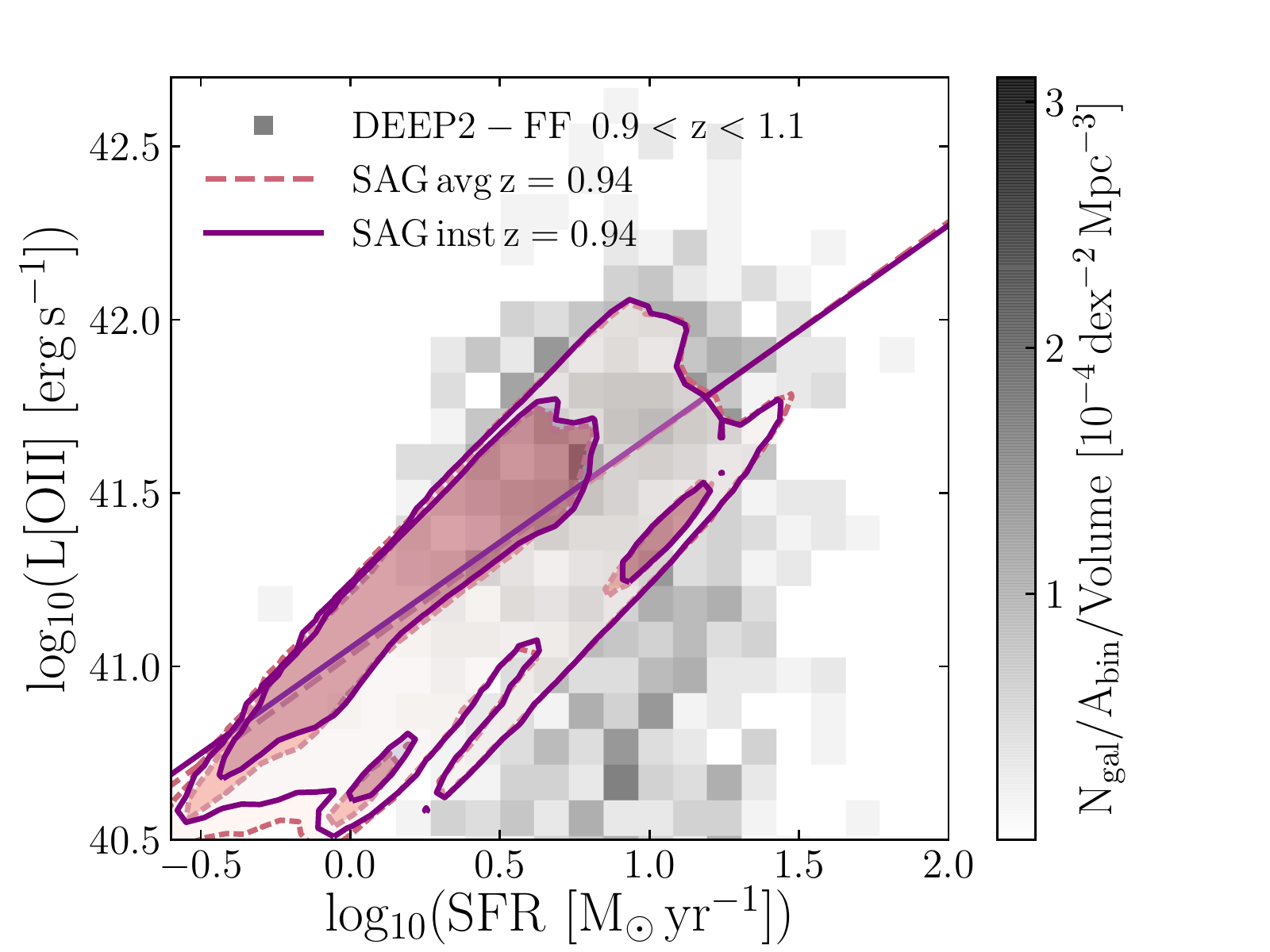}\vspace{-0.3cm}
\caption{Intrinsic \OII luminosity as a function of the SFR for the \SAG model galaxies at $z\sim1$ (contours) and the DEEP2-FF observations at $0.9<z<1.1$ (grey, shaded squares). The bar represents the number density of DEEP2-FF galaxies in each 2D bin normalised by the bin area in units of [dex$^{-2}$\,Mpc$^{-3}$]. We have imposed a minimum \OII flux of 5$\times 10^{-18}$erg\,s$^{-1}$\,cm$^{-2}$ to both observations and models. The model \LOII values are calculated by assuming instantaneous (solid, purple contours) and average (dashed, salmon) SFR as input for the \GE prescription. The innermost (outermost) model contours encompass 68\% (95\%) percent of the galaxy distributions. The diagonal lines represent the \LOII-SFR correlations, whose coefficients are given in Table\,\ref{tab:params_prop_SAG}.}
\label{fig:SFRinstavg}
\end{center}
\end{figure}

Fig.\,\ref{fig:SFRinstavg} shows, as a function of SFR, the intrinsic (i.e. corrected from dust attenuation) \LOII that the coupling with \GE gives for both the instantaneous (solid contours) and average (dashed) SFR from \SAG at $z\sim1$. The innermost (outermost) contours enclose 68\% (95\%) of our model galaxies. The diagonal lines show the correlations between SFR and \LOII. These are tight correlations, whose best-fitting parameters are reported in Table\,\ref{tab:params_prop_SAG}. Under laid are the DEEP2-FF observational data at $0.9<z<1.1$. Overall, the model galaxy distributions presented in Fig.\,\ref{fig:SFRinstavg} are very similar for the \LOII derived from either the instantaneous or the average SFRs. These distributions show a bimodality that can also be seen in the observations.

The instantaneous and average SFR derived distributions differ the most at SFR$\lesssim10^0\,\rm{yr^{-1}M_{\odot}}$, with \OII luminosities from average SFR being $\sim0.2$~dex fainter than those from instantaneous SFR. At SFR$\sim10^{1.5}$yr$^{-1}$M$_{\odot}$, there are slightly less bright \OII emitters from instantaneous SFR. 

DEEP2-FF galaxies in the upper density peak of the observed bimodal distribution shown in  Fig.\,\ref{fig:SFRinstavg} are older, more massive, more luminous and slightly more star-forming (mean values: $\langle {\rm age}\rangle\sim10^{9.28}$yr, $\langle$\Mstar$\rangle\sim10^{10.42}$M$_{\odot}$, $\langle$\LOII$\rangle\sim10^{41.48}$erg$\,$s$^{-1}$, $\langle$SFR$\rangle\sim10^{1.13}$yr$^{-1}$M$_{\odot}$) compared to their counterparts in the lower density area ($\sim10^{9.13}$yr, $\sim10^{10.22}$M$_{\odot}$, $\sim10^{39.07}$erg$\,$s$^{-1}$, $\sim10^{1.08}$yr$^{-1}$M$_{\odot}$). Overall, we find an opposite trend for model galaxies. In fact, the upper peak of the bimodality is composed of younger, less massive, slightly more luminous, less star-forming galaxies with mean values: $\langle$age$\rangle\sim10^{9.16}$yr, $\langle$\Mstar$\rangle\sim10^{9.58}$M$_{\odot}$, $\langle$\LOII$\rangle\sim10^{41.26}$erg$\,$s$^{-1}$, $\langle$SFR$\rangle\sim10^{0.21}$yr$^{-1}$M$_{\odot}$); the lower peak has mean values: $\langle$age$\rangle\sim10^{9.32}$yr, $\langle$\Mstar$\rangle\sim10^{10.06}$M$_{\odot}$, $\langle$\LOII$\rangle\sim10^{41.25}$erg$\,$s$^{-1}$, $\langle$SFR$\rangle\sim10^{0.83}$yr$^{-1}$M$_{\odot}$.

\begin{figure}
\begin{center}
\includegraphics[width=1.05\linewidth]{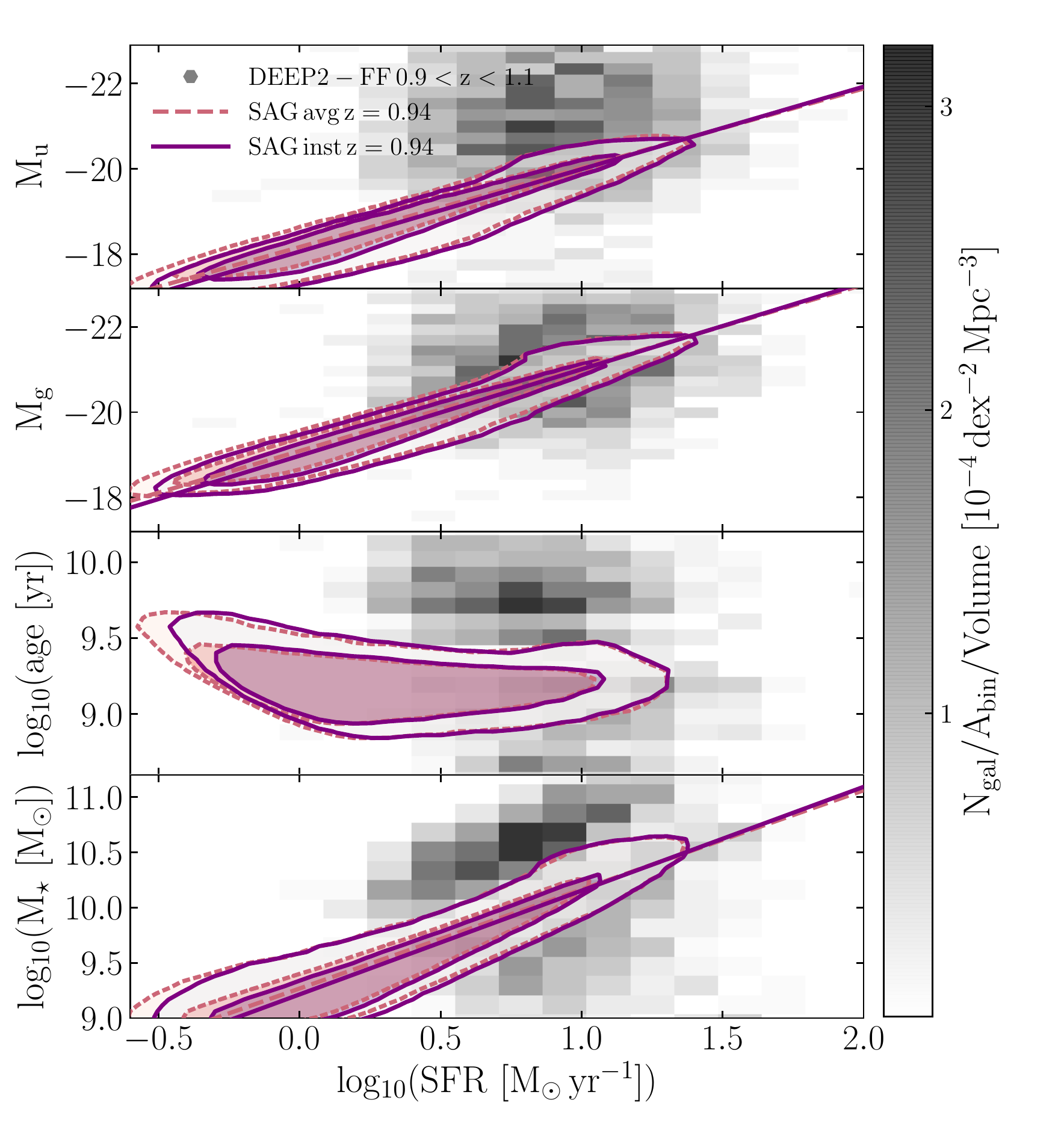}\vspace{-0.6cm}
\caption{From top to bottom: intrinsic magnitudes, ages and stellar masses as a function of star formation rate for \SAG (contours) at $z\sim1$ and  DEEP2-FF observations at $0.9<z<1.1$ (grey, shaded squares). The colour bar shows the number density of DEEP2 galaxies per bin area in units of [dex$^{-2}$\,Mpc$^{-3}$]. The dashed, salmon (solid, purple) contours represent the average (instantaneous) SFRs. The innermost (outermost) contours encompass 68\% (95\%) of the distributions. The diagonal lines are the linear fits showing the significant correlations (i.e. $r\ge0.6$), whose coefficients are reported in Table\,\ref{tab:params}, together with the best-fit parameters.}
\label{fig:proponly}
\end{center}
\end{figure}

At the end of this Section, we will discuss further the origin of the DEEP2-FF \LOII-SFR bimodal trend in connection with other galactic properties shown in Fig.~\ref{fig:proponly}.

\begin{figure}\vspace{-0.2cm}
\begin{center}
\includegraphics[width=\linewidth]{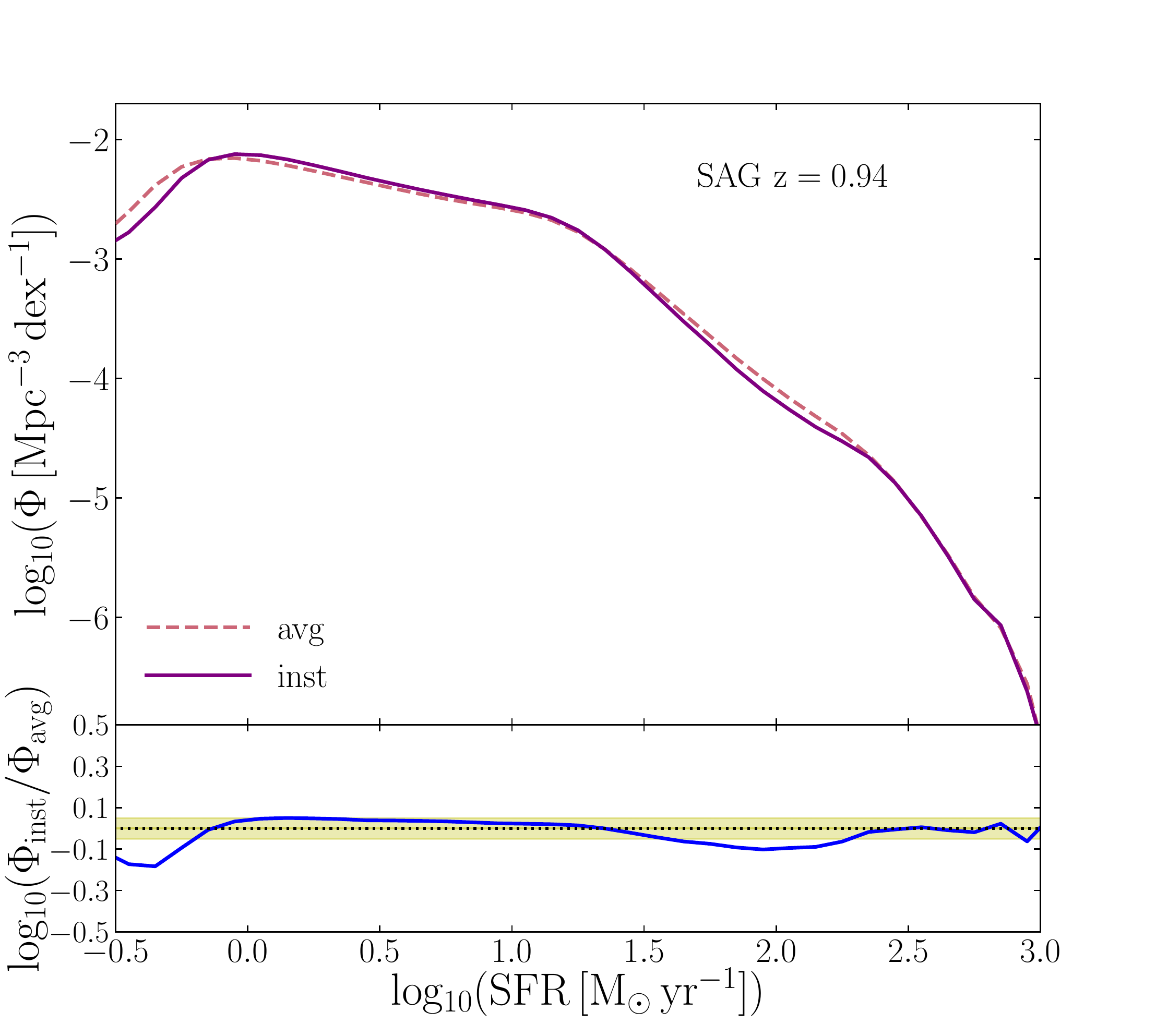}\vspace{-0.4cm}
\caption{Average (dashed, salmon) versus instantaneous (solid, purple) SFR functions for SAG model galaxies. The bottom panel shows the ratio between the two, and the yellow, shaded region highlights the 5\% region of agreement.}
\label{fig:ratioSFRfunc}
\end{center}
\end{figure}
\begin{figure}\vspace{-0.5cm}
\begin{center}
\includegraphics[width=\linewidth]{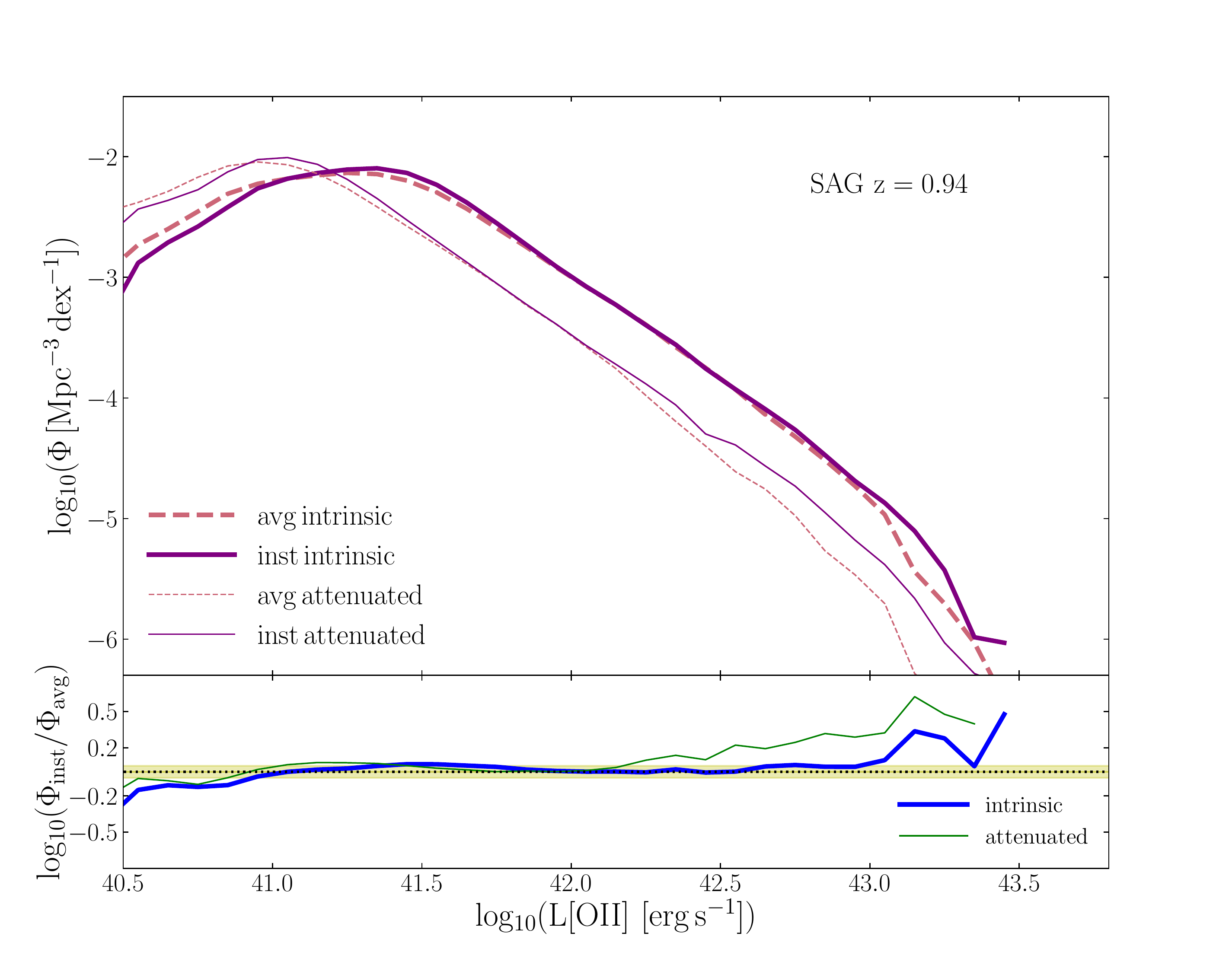}\vspace{-0.4cm}
\caption{Intrinsic (thick lines) and attenuated (thin) \OII luminosity functions based on \SAG average (dashed, salmon) and instantaneous SFR (solid, purple). The bottom panel shows the ratios between the two and the yellow stripe highlights the 5\% region of agreement. We apply the mocks the same \OII flux limit of DEEP2-FF observations, $5\times 10^{-18}$erg\,s$^{-1}$\,cm$^{-2}$ (see Sec.\,\ref{sec:samplefinale}).}
\label{fig:ratioLFfunc}
\end{center}
\end{figure}

In the top panel of Fig.\,\ref{fig:ratioSFRfunc} we compare the average (dashed, salmon) and instantaneous (solid, purple) \SAG SFR functions at $z\sim1$, whose ratio is displayed in the bottom panel. The instantaneous and average SFR functions remain within 5\% of each other at SFR$>10^0\,\rm{yr^{-1}M_{\odot}}$ (the 5\% region is highlighted by the yellow shade). There is a slightly larger fraction, within 20\%, of SAG galaxies having low average SFR, SFR$<10^0\,\rm{yr^{-1}M_{\odot}}$, than instantaneous values. The main difference between average and instantaneous SFRs is found for galaxies with the highest specific SFR (i.e., SFR/\Mstar) and stellar masses below $10^{11}$M$_\odot$.

The top panel in Fig.\,\ref{fig:ratioLFfunc} presents the intrinsic (thick lines) and attenuated (thin) \OII luminosity functions derived from the average SFR (dashed, salmon line) and instantaneous SFR (solid, purple) from SAG. We impose on the \SAG model galaxies the same \OII flux limit of DEEP2-FF observations, $5\times 10^{-18}$erg\,s$^{-1}$\,cm$^{-2}$ (see Sec.\,\ref{sec:samplefinale}), which corresponds to \LOII$\sim10^{40.4}\,$erg\,s$^{-1}$ at $z=1$ in Planck cosmology \citep[][]{Planck15}. 
The instantaneous-to-average amplitude ratios are displayed in the bottom panel of Fig.\,\ref{fig:ratioLFfunc}. The intrinsic (attenuated) \LOII functions have differences below 5\% for luminosities in the range $10^{41}-10^{43}\,$erg\,s$^{-1}$ ($10^{41}-10^{42.2}\,$erg\,s$^{-1}$), which are highlighted by the yellow shade. At lower (higher) luminosities, the discrepancies grow up to 20\% (30\%). For the brightest galaxies, the discrepancy remains within 50\%. The difference produced in \LOII by assuming average instead of instantaneous SFR does not change significantly with redshift over the range $0.6<z<1.2$ (see Appendix\,\ref{sec:appendix1} for further details). Thus, the average and instantaneous SFR can be assumed interchangeably for average galaxies.

\begin{table}
\centering
\begin{tabular}{@{}lccccc@{}}\toprule
y=$A\,x+B$&  $A$&$B$&$\sigma_{\rm y}$&$r$ \\
\midrule
y=log$_{10}$(\LOII)&&&&\\
x=log$_{10}({\rm SFR}_{\rm avg})$&0.625$\pm$0.001&41.03$\pm$0.01&0.40&0.83\\
x=log$_{10}({\rm SFR}_{\rm inst})$&0.609$\pm$0.001&41.05$\pm$0.01&0.38&0.80\\
\midrule
y=\magu&&&&\\
x=log$_{10}({\rm SFR}_{\rm avg})$&-1.859$\pm$0.001&-18.17$\pm$0.01&1.07&0.92\\
x=log$_{10}({\rm SFR}_{\rm inst})$&-1.934$\pm$0.001&-18.06$\pm$0.01&1.07&0.90\\
\midrule
y=\magg&&&&\\
x=log$_{10}({\rm SFR}_{\rm avg})$&-1.951$\pm$0.001&-19.09$\pm$0.01&1.11&0.93\\
x=log$_{10}({\rm SFR}_{\rm inst})$&-2.029$\pm$0.001&-18.98$\pm$0.01&1.11&0.91\\
\midrule
y=log$_{10}($M$_{\star})$&&&&\\
x=log$_{10}({\rm SFR}_{\rm avg})$&0.897$\pm$0.001&9.27$\pm$0.01&0.54&0.89\\
x=log$_{10}({\rm SFR}_{\rm inst})$&0.939$\pm$0.001&9.21$\pm$0.01&0.54&0.87\\
\bottomrule
\end{tabular}
\caption{Best-fit parameters of the linear scaling relations found for \SAG model galaxies at $z=1$ and shown in Fig.\,\ref{fig:proponly}. The parameter $r$ is the correlation coefficient and $\sigma_y$ is the scatter in the $y$-axis. All the \LOII values are intrinsic.}
\label{tab:params_prop_SAG}
\end{table} 

In Fig.\,\ref{fig:proponly}, from top to bottom, we display the \SAG broad-band $u$ and $g$ absolute magnitudes, ages and stellar masses as a function of the average SFR (dashed, salmon contour) and instantaneous SFR (solid, purple). We compare them with the DEEP2-FF observations at $0.9<z<11$ (grey, shaded squares). Except for the age, all these properties are tightly correlated with both SFRs. The lack of correlation between age and SFRs is clear for both the model and DEEP2-FF galaxies. We fit straight lines to the instantaneous and average contours and report the best-fit parameters and correlation coefficients in Table\,\ref{tab:params_prop_SAG}. For the broad-band magnitudes, the slopes of the average SFR correlations are only $\sim0.07$ shallower than the instantaneous ones; for the stellar mass they are even closer.
Overall, the width of the distributions as a function of both SFRs does not vary significantly. The average SFR contours extend down to slightly smaller values compared to the instantaneous contours.

The DEEP2-FF age and stellar mass distributions as a function of SFR in Fig.\,\ref{fig:proponly} show a bimodal trend, with an upper population of older, more massive, luminous, quiescent galaxies  ($\langle \rm{age}\rangle\sim10^{9.8}\,$yr, $\langle M_{\star}\rangle\sim10^{10.82}\,{\rm M}_{\odot}$, $\langle$\LOII$\rangle\sim10^{40.57}\,{\rm erg}\,{\rm s}^{-1}$, $\langle\rm{SFR}\rangle\sim10^{1.02}\,{\rm yr}^{-1}{\rm M}_{\odot}$) and a lower tail of younger, less massive, luminous, more star-forming emitters ($\langle \rm{age}\rangle\sim10^{8.48}\,$yr, $\langle M_{\star}\rangle\sim10^{9.72}\,{\rm M}_{\odot}$, $\langle$\LOII$\rangle\sim10^{40.45}\,{\rm erg}\,{\rm s}^{-1}$, $\langle\rm{SFR}\rangle\sim10^{1.24}\,{\rm yr}^{-1}{\rm M}_{\odot}$). We obtain the same mean galaxy properties splitting the DEEP2-FF sample with a cut in either the age-SFR or Mstar-SFR planes.

The mean DEEP2-FF values derived from splitting in age or stellar mass as a function of SFR are similar to those obtained by splitting in \LOII versus SFR (see Fig.\,\ref{fig:SFRinstavg}).
The age/mass-SFR bimodal trend observed in DEEP2-FF galaxies is not reproduced by the SAG model galaxies, which instead look bimodal in the \LOII-SFR plane (Fig.\,\ref{fig:SFRinstavg}) because of the non-trivial dependence of \LOII on metallicity through the parameters $q_0$ and $\gamma$ (see Sec.\,\ref{sec:O2sams}).

In this section, we have shown that using the SAG average SFRs as input for the \GE code gives results within 5\% from using the instantaneous value for galaxies with attenuated \LOII in the range $10^{40.9}-10^{42.2}\,$erg\,s$^{-1}$, and with intrinsic \LOII between $10^{40.9}-10^{43}\,$erg\,s$^{-1}$. These are the ELGs with SFR within $10^{-0.2}-10^{1.6}\,$yr$^{-1}$\,M$_{\odot}$. At higher and lower SFRs, there is a larger discrepancy between the average and instantaneous values, which translates into a larger difference ($<60\%$) in the number of bright \OII emitters. Thus, this effect is not significant for the average galaxy population.

\subsection{Model {\oiitit} luminosity functions}
\label{sec:LFsams}
\begin{figure}
\centering
\includegraphics[width=\columnwidth]{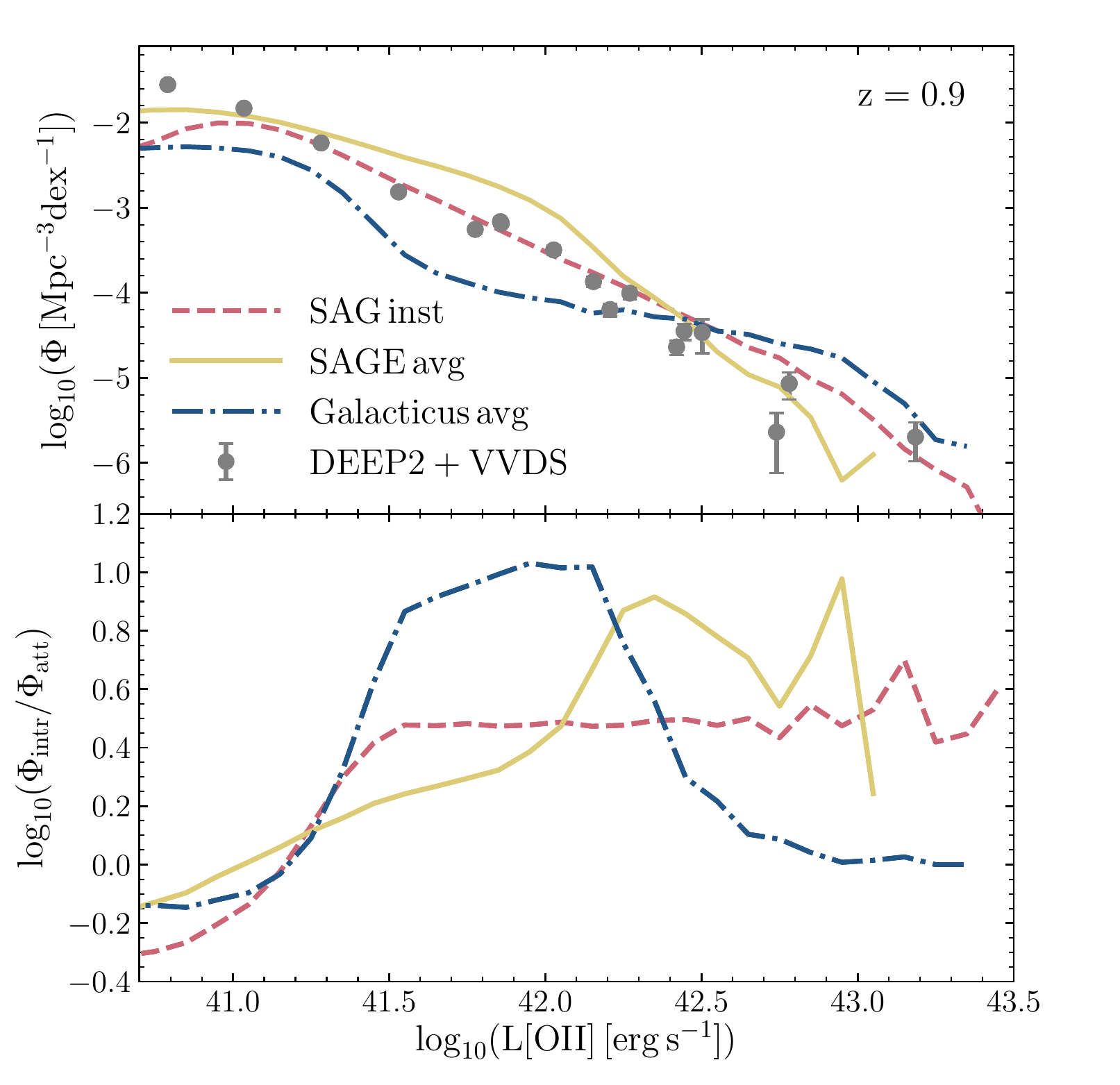}\vspace{-0.5cm}
\caption{\textit{Top:} Dust attenuated \OII luminosity functions of the \MD at $z\sim1$ compared with DEEP2+VVDS observations \citep[]{Comparat2016LFs}. We consider all SAM galaxies above $5\times10^{-18}$\,erg\,s$^{-1}$\,cm$^{-2}$. All the \OII luminosities are computed using the \GE code with SFR and cold gas metallicity as inputs (see Section\,\ref{sec:O2sams}). The SAG \LOII, which are estimated using the instantaneous SFR, are in good agreement with the SAGE and Galacticus results based on the average SFR.  \textit{Bottom:} Ratios between the model intrinsic \OII luminosity functions (given by Eq.\,\ref{eq:lum}) and the dust attenuated ones (see Sec.\,\ref{sec:dust}) shown in the upper panel.}
\label{fig:LFz}
\end{figure}

In the top panel of Fig.~\ref{fig:LFz}, we present the \MD dust attenuated \OII luminosity functions at $z=0.94$ compared to a compilation of DEEP2 and VVDS data from \citet{Comparat2016LFs}. Note that similar results have been found within the redshift range $0.6<z<1.2$, although they are not shown here. The SAM \OII luminosities have been derived using the \GE code described above coupled with instantaneous SFR for SAG model galaxies, and average SFR for SAGE and Galacticus, for which the instantaneous quantity is not available. The dust attenuation has been accounted for by correcting these luminosities applying Eq.~\ref{eq:attenuation} with \citet{Cardelli1989} extinction curve. 
There are varying degrees of agreement between the models and observational data across the $\sim$3 decades in \OII luminosity and redshift range considered. Nevertheless, the trends from all the data sources are consistent. This plot highlights that the shape and normalisation of a predicted \OII luminosity function from a SAM are robust to both the precise prescriptions that govern galaxy evolution in the model, and the calculation of \OII from either instantaneous or average SFR. 

In the top panel of Fig.~\ref{fig:LFz}, we see a drop in the number of Galacticus \OII emitters at intermediate luminosities that is independent of the stellar mass. This is mainly determined by the half mass radii of the disc, R$_{1/2}^{\rm disc}$, that enter the dust attenuation correction (see Eq.~\ref{eq:columndens}). These radii in Galacticus are about 50\% smaller than in SAG and SAGE.  
At \LOII$\gtrsim10^{42.5}$erg~s$^{-1}$ and $z<1.2$, Galacticus predicts about 0.5~dex more \OII emitters than the other two models.

In the bottom panel of Fig.\,\ref{fig:LFz}, we display the ratios of the attenuated-to-intrinsic \LOII functions. As expected, the largest effect of attenuation occurs at \LOII$\gtrsim10^{42}\,$erg\,s$^{-1}$, where more massive galaxies are located, while in the low-luminosity, low-mass regime, the observed and intrinsic signals tend to overlap. The ratio between the intrinsic and attenuated luminosity functions is model dependent. In particular, the largest variations are due to the dust model, which depends on the metallicity, gas content and size of each galaxy, as described in \S\,\ref{sec:dust}. For \SAGE model galaxies, this ratio increases for brighter \LOII galaxies. For \SAG, the ratio also increases up to \LOII$\gtrsim10^{41.5}\,$erg\,s$^{-1}$, although with a steeper slope, and beyond this value it reaches a plateau. The ratio for \GAL has a prominent bump in the luminosity range $10^{41.5}-10^{42.5}\,$erg\,s$^{-1}$, where the effect of attenuation is more pronounced, and this feature corresponds to the drop seen at intermediate \LOII in the upper panel. At higher luminosities, there is almost no difference between the intrinsic and dust attenuated \GAL \LOII functions.

\section{{\oiitit} luminosity proxies}
\label{sec:proxies}
Observational studies have shown tight correlations between the \OII luminosity, SFR \citep[]{1998ARA&A..36..189K, 2012MNRAS.420.1926S, kewley04, 2006ApJ...642..775M, comparat2015} and the galaxy UV-emission \citep[]{comparat2015}, \textit{without} the need to introduce any dependence on metallicity \citep[]{2006ApJ...642..775M}. 
This has prompted authors of theoretical papers to treat star-forming galaxies as ELGs when making predictions for upcoming surveys \citep[e.g.][]{2018orsi, 2019arXiv190604298J}. 

Here we explore the possibility of using simple, linear relations to infer the \OII luminosity from global galaxy properties that are commonly output in SAMs. For this purpose, we investigate both observationally motivated prescriptions (Section\,\ref{sec:SFR-LO2}), and we derive model relations from the \GE code coupled with the SAMs considered (Sections\,\ref{sec:LO2_mags} and \ref{sec:LO2_age_Z}). For this last study, we quantify the correlation between the model \LOII from \GE with the average SFR, broad-band magnitudes, stellar masses, ages and cold gas metallicities. Directly using the measured \LOII-SFR linear relation is useful to understand when is adequate to consider ELGs equivalent to star-forming galaxies and when it is not.

We find that the stellar mass of the \MD are unaffected by the change in proxies for estimating their \OII luminosities. As a consequence, the stellar-to-halo mass relation (SHMR) is also unchanged using different \LOII proxies.

We remind the reader that, unless otherwise specified, we exclusively select emission line galaxies with fluxes above $5\times10^{-18}\rm{erg\,s^{-1}\,cm^{-2}}$ in both the DEEP2-FF observations and \textsc{MultiDark-Galaxies}. This flux limit corresponds to a \LOII$>10^{40.4}\,$erg\,s$^{-1}$ at $z=1$ in the Planck cosmology \citep[]{Planck15}. All the results in what follows have these minimum cuts applied.

\subsection{The SFR--L{\oiitit} relation}
\label{sec:SFR-LO2}

In this Section, we derive {\bf intrinsic} \LOII from the average SFR of the \MD using three different, published relations assuming a \citet[][]{1998ARA&A..36..189K} IMF. These are: the \citet[][]{2006ApJ...642..775M} conversion
\citep[see also][]{comparat2015} calibrated at $z=0.1$,
\begin{equation}
\rm{L_{[OII]}^{Moust}}({\rm{erg}\,\,s}^{-1})=\frac{SFR({\rm{M}}_{\odot}\,{\rm{yr}}^{-1})}{2.18\times 10^{-41}},
\label{eq:LA}
\end{equation}
the \citet[][]{2012MNRAS.420.1926S} formulation optimised at $z=1.47$,
\begin{equation}
\rm{L_{[OII]}^{Sob}}({\rm{erg}\,\,s}^{-1})=\frac{SFR({\rm{M}}_{\odot}\,{\rm{yr}}^{-1})}{1.4\times 10^{-41}},
\label{eq:LB}
\end{equation}
the \citet[][]{kewley04} conversion calibrated at $z=1$,
\begin{equation}
\begin{aligned}
\rm{L_{[OII]}^{Kew}}(erg\,s^{-1})=\frac{SFR({\rm{M}}_{\odot}\,{\rm{yr}}^{-1})}{7.9\times 10^{-42}}\\
\times (a[12+\log_{10}({\rm{O/H}})_{\rm{cold}}]+b).
\end{aligned}
\label{eq:LC}
\end{equation}
The coefficients $(a,b)$ in the equation above are the values from \citet[][]{kewley04}
derived for the $R_{23}$ metallicity diagnostic \citep[]{pagel1979}. The $[12+\log_{10}(\rm{O/H})_{\rm{cold}}]$ term is the
\OII ELG gas-phase oxygen abundance, which we proxy with the cold gas-phase metallicity $Z_{\rm{cold}}$ given in Eq.\,\ref{eq:metallicity} through the solar abundance and metallicity.
Explicitly we have: 
\begin{equation}
{12+\log_{10}{\rm{(O/H)_{cold}}}}=[{12+\log_{10}{\rm{(O/H)_{\odot}}}}]\frac{Z_{\rm{cold}}}{Z_{\odot}},
\label{eq:gal}
\end{equation}
where we assume $Z_{\odot}=0.0134$ \citep[][]{2009ARA&A..47..481A}, and $[12+\log_{10}{\rm (O/H)_{\odot}}] =8.69$ \citep[][]{2001ApJ...556L..63A}. {As the above relations are for intrinsic luminosities, dust attenuated quantities are obtained following the description in \S~\ref{sec:dust}.}

For \SAG and \GAL, galaxies' cold gas is broken into bulge and disc components (see their respective papers for their definitions of a `gas bulge'); we therefore take a mass-weighted average of these components' metallicities to obtain $Z_{\rm cold}$. \SAGE instead always treats cold gas as being in a disc. In addition, the \SAG catalogues also output the $\rm{(O/H)_{cold}}$ values, which are mass density ratios, that we use in the calculation of Eq.\,\ref{eq:LC} for \SAG model galaxies. In order to derive the correct abundances in terms of number densities, we need to rescale them
by the Oxygen-to-Hydrogen atomic weight ratio, $A_{\rm{O}}/A_{\rm{H}}\sim15.87$.

\begin{figure}
\begin{center}
\includegraphics[width=7cm]{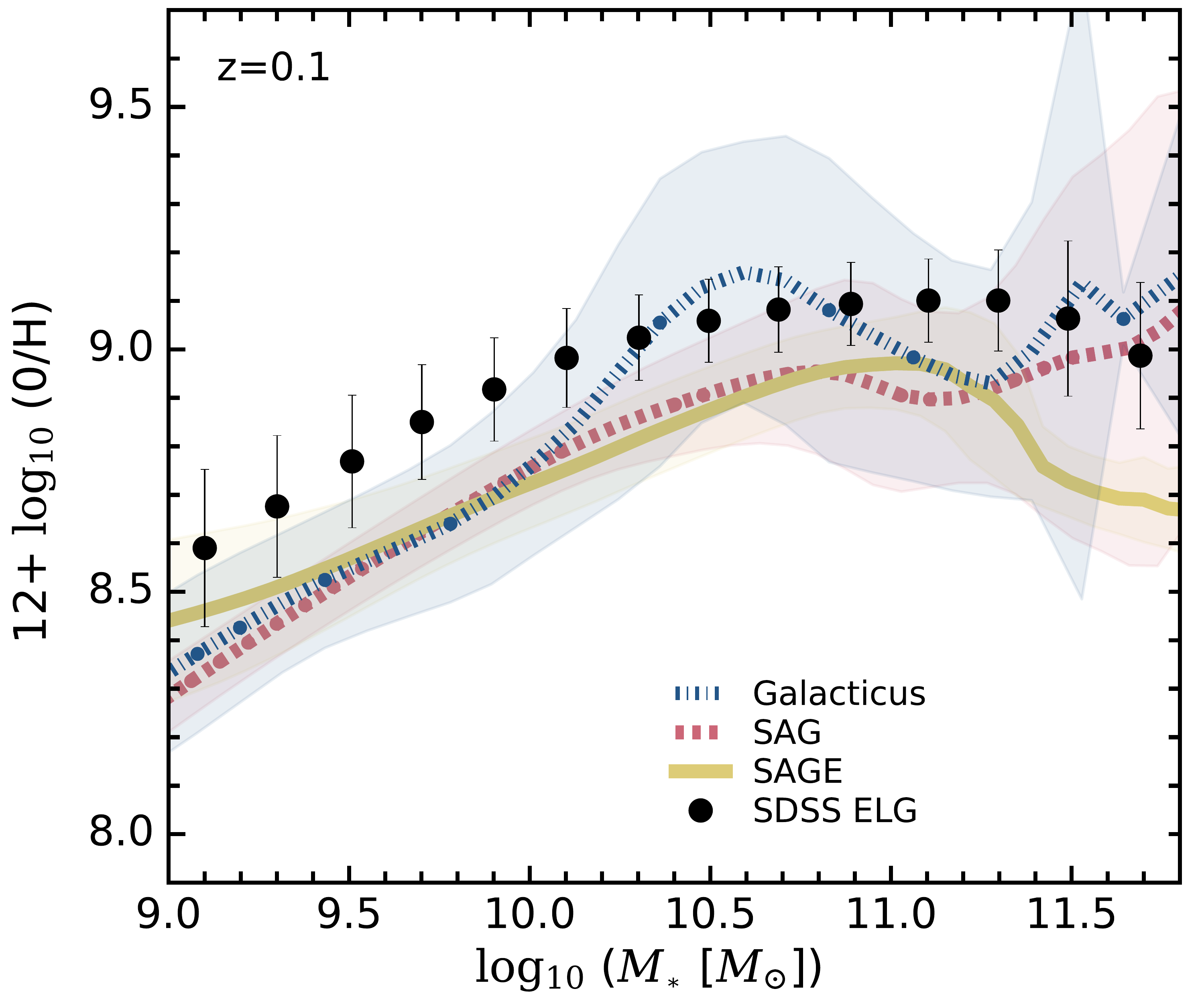}
\caption{Mean gas-phase oxygen abundance in bins of stellar mass of the SDSS emission
line galaxies at $z\sim0.1$ \citep[][]{2017MNRAS.472..550F} compared to the \MD models.
The abundance is computed for the SAMs using Eq.~\ref{eq:gal}. The error bars on the SDSS
measurements are the $1\sigma$ scatter around the mean.}
\label{fig:metall_vs_sm2}
\end{center}
\end{figure}
Fig.\,\ref{fig:metall_vs_sm2} displays the comparison between the gas-phase oxygen abundances of our SAM galaxies computed using Eq.\,\ref{eq:gal} and the observed abundance of the SDSS \OII ELGs at $z\sim0.1$ from \citet[][]{2017MNRAS.472..550F}. The SDSS metallicity values have been derived from the MPA-JHU DR7\footnote{\label{note1}\url{http://wwwmpa.mpa-garching.mpg.de/SDSS/DR7/}} catalogue of spectrum measurements and are built according to the works of \citet[][]{Tremonti2004} and \citet[][]{Brinchmann2004}. Overall, we find that the gas-phase oxygen abundance in \MD increases with stellar mass up to \Mstar$\sim10^{11}\,$M$_{\odot}$. Beyond this value it drops and reaches a plateau. 

The \SAG and \SAGE model galaxies under-predict the gas-phase oxygen abundance by an average factor of $\sim0.02\,$dex. This systematic offset for SAGE is not predictive, but purely due to the fact that this model was calibrated by assuming a different value of ${\rm{(O/H)}}_{\odot}/Z_{\odot}$, specifically ${\rm{[12+log_{10}(O/H)] = [9+log_{10}}}(Z_{\rm{cold}}/0.02)]$; for further details, see \citet[][]{knebe2018}.  

At \Mstar$<10^{10.2}\,$M$_{\odot}$, \GAL also under-predicts the gas-phase abundance by the same factor. However, this model exhibits a bump at \Mstar$\sim10^{10.5}\,$M$_{\odot}$. 
This feature is related to the excess of galaxies around this stellar mass, which is seen in the galaxy stellar mass function (see Fig.\,\ref{fig:massf}). This excess was found to be produced by the depletion of gas due to the extreme AGN feedback mechanism implemented in \GAL, where the galaxies have almost no inflow of pristine gas, and rapidly consume their gas supply \citep[for further details, see][]{knebe2018}. 

We have investigated further this feature finding that, if we exclude galaxies with progressively higher cold gas fraction (CGF), which is defined as CGF=$M_{\rm{cold}}$/$M_\star$, the bump shrinks continuously. Fig.\,\ref{fig:metall_vs_sm2} is produced by combining two cuts: CGF$>0.1$ and sSFR$>10^{-11}\,$yr$^{-1}$. The first one eliminates about half of the \GAL model galaxies, most of them with unrealistically small CGFs, possibly meaning that their metallicities are not reliable due to the precision used in evolving the relevant ordinary differential equations \citep[][]{2012NewA...17..175B}. 
The second cut selects only very star-forming galaxies. The bump completely disappears for CGF$>0.5$, but in that case $\sim70\%$ of the galaxies are excluded from the sample.
\begin{figure*}
\begin{center}
\includegraphics[width=1.1\linewidth]{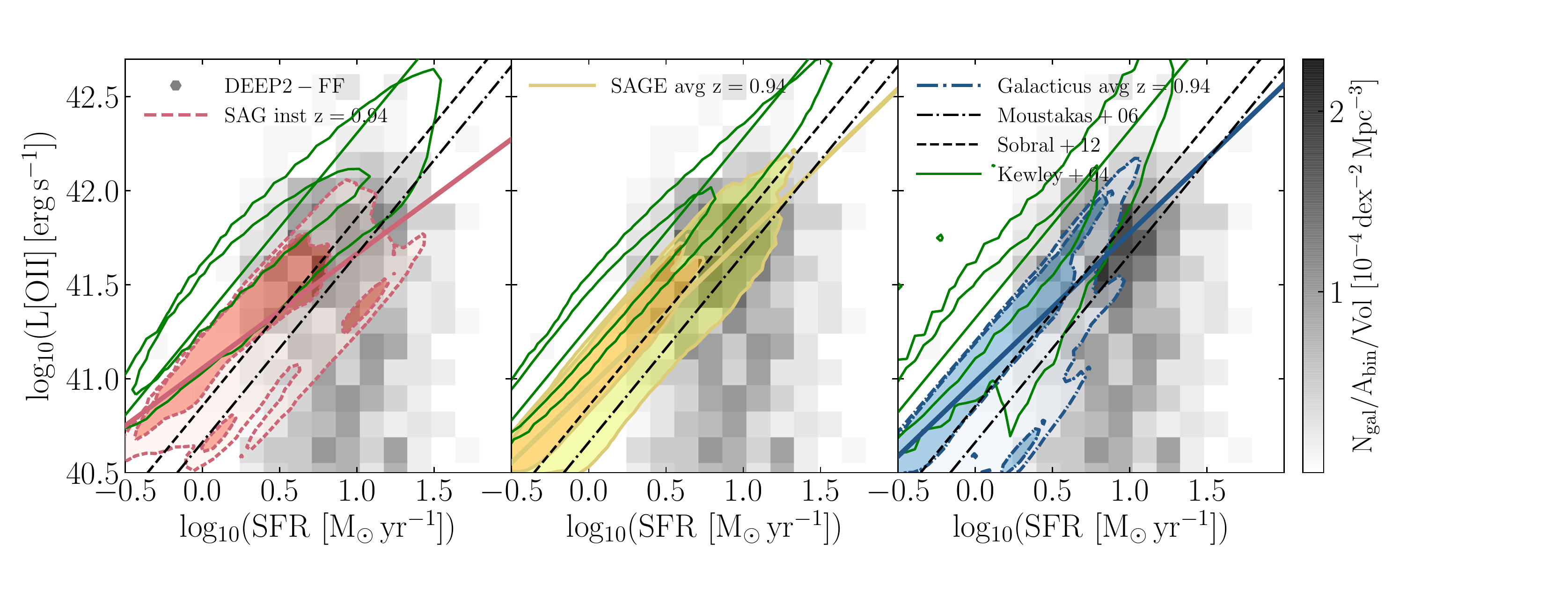}\vspace{-0.8cm}
\caption{Intrinsic \OII luminosity as a function of the SFR from the \MD at $z\sim1$ (salmon, yellow and blue, filled contours), compared with the DEEP2-FF observations at $0.9<z<1.1$ (grey, shaded squares, colour-coded with the density of emitters per 2D bin area). The innermost (outermost) contour represents 68\% (95\%) of the galaxy distributions. For \SAG model galaxies, the \OII luminosities have been computed from instantaneous SFRs, while for the other SAMs they are based on average SFRs. Both data and model ELGs are selected imposing a minimum \OII flux of 5$\times10^{-18}$ erg s$^{-1}$ cm$^{-2}$. The thick, coloured, diagonal lines are the linear fits to each SAM distribution, and their best-fit parameters are reported in Table\,\ref{tab:params}. The black dot-dashed and dashed, diagonal lines are the \LOII predictions obtained from the SFR range of interest using Eqs.\,\ref{eq:LA} and \ref{eq:LB}, respectively. The green, empty contours are the \citet[][]{kewley04} predictions obtained using Eq.\,\ref{eq:LC} with SFR and cold gas metallicity as inputs. The green, solid lines are the same predictions assuming median metallicity values in bins of SFR.}
\label{fig:nz}
\end{center}
\end{figure*}

Fig.\,\ref{fig:nz} compares  the intrinsic \OII luminosity as a function of SFR for the three SAMs (coloured, filled contours) with the DEEP2-FF data at $z\sim1$ (grey, shaded squares). We also show the results of the conversions given in Eqs.\,\ref{eq:LA}-\ref{eq:LC} (diagonal, black and green lines).
The model \LOII is computed using the \GE code coupled with instantaneous SFR for SAG, and average SFR for the other semi-analytic models. The distributions of \SAG, \SAGE and \GAL behave in a similar way, reproducing the bimodality observed in the data. The coloured lines (dashed, salmon; solid, yellow; dot-dashed, blue) are the linear fits to the model \LOII-SFR correlations. The best-fit parameters, correlation coefficients ($r$-values) and dispersions in both directions are reported in Table\,\ref{tab:params}.

Fig.\,\ref{fig:nz} shows that all the model galaxies considered overlap with the DEEP2-FF observations and extend further towards lower SFR values. All three SAMs cover the \LOII observational range with their $2\sigma$ regions. \SAGE and \GAL get to the very bright domain of the parameter space, while \SAG is limited to fainter \LOII values.

All the SAMs are tightly correlated in the SFR--luminosity plane and such a trend is in reasonable agreement with the observationally derived relations from Eqs.\,\ref{eq:LA}-\ref{eq:LC} (diagonal, black and green lines). 

In Fig.\,\ref{fig:nz}, the \citet[]{kewley04} parametrisation (green line and contours in Fig.\,\ref{fig:nz}) appears above all the \GE derivations. These contours are obtained from Eq.\,\ref{eq:LC}, by inputting instantaneous (average) SFR and cold gas metallicity for \SAG (\SAGE, \GAL) model galaxies. The green, straight lines are calculated by feeding the median metallicity values in bins of SFR into Eq.\,\ref{eq:LC}. Although both the \citet[]{kewley04} relation and the \GE code assume the same cold gas metallicity values as inputs, the obtained distributions are very different. The width of the distributions is model-dependent and the \LOII obtained for galaxies in \SAG and Galacticus present bimodal distributions. This bimodality comes from the MAPPINGS-III term $F(\lambda_j, q, Z_{\rm{cold}})$ in Eq.\,\ref{eq:lum}, that is a non-linear function of $Z_{\rm{cold}}$. 
\begin{figure*}
\begin{center}\vspace{-0.3cm}\hspace{-1cm}
\includegraphics[width=1.1\linewidth]{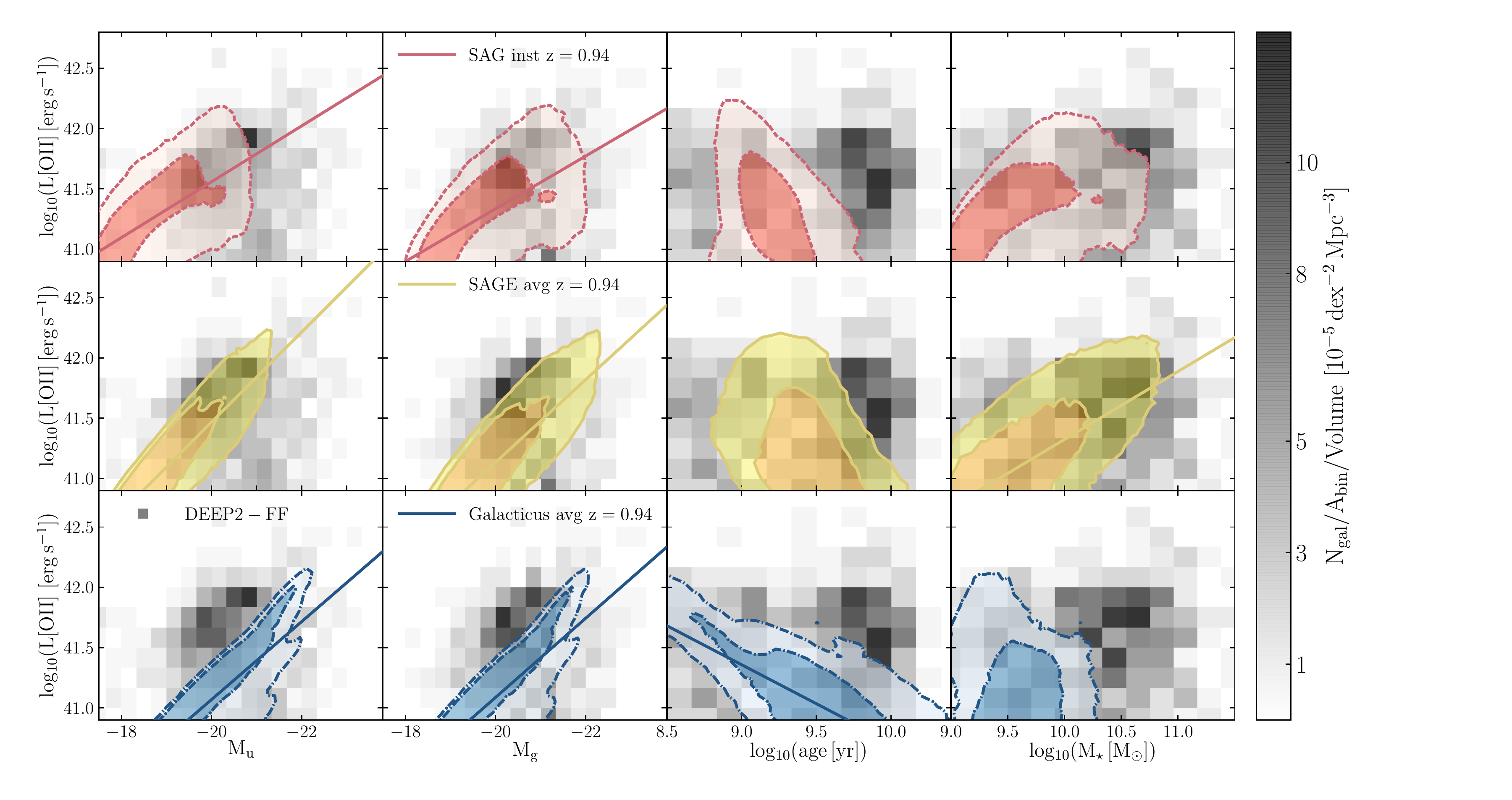}\vspace{-0.8cm}
\caption{From top to bottom and from left to right: \SAG, \SAGE and \GAL $z\sim1$ intrinsic \OII luminosities versus broad-band magnitudes, ages and stellar masses (contours) compared with the  DEEP2-FF observations at $0.9<z<1.1$ (grey, shaded  squares). The \LOII values are computed using the \GE code with instantaneous SFR for \SAG and average SFR for \SAGE and \GAL. The innermost and outermost model contours represent 68\% ($1\sigma$) and 95\% ($2\sigma$) of the distribution. A minimum \OII flux cut of $\rm{5\times10^{-18}\,\,erg\,s^{-1}\,cm^{-2}}$ has been applied to both data and model galaxies. The diagonal lines are the linear fits for strong correlations with $r>0.6$, as reported in Table\,\ref{tab:params}.} 
\label{fig:propall}
\end{center}
\end{figure*}

\begin{table*}\centering
\begin{tabular}{@{}llcccc@{}}\toprule
\hspace{3cm}z=1&  &\SAG& \SAGE& \GAL  \\
\midrule
\logaLOII\,=\,$A$\,log$_{10}$(SFR/M$_{\odot}$\,yr$^{-1}$)+$B$&$A$&0.609$\pm$0.001&0.792$\pm$0.001&0.795$\pm$0.001&\\
&$B$&41.05$\pm$0.01&40.98$\pm$0.01&40.95$\pm$0.01\\
&$\sigma_{\rm{\log(SFR)}}$&0.50&0.53&0.48\\
&$\sigma_{\rm{\log(L[OII])}}$&0.38&0.45&0.46\\
&$r$&0.80&0.92&0.83\\
\midrule
\logaLOII\,=\,$A$\,\magu+$B$&$A$&-0.231$\pm$0.001&-0.373$\pm$0.001&-0.323$\pm$0.001&\\
&$B$&36.93$\pm$0.01&34.01$\pm$0.01&34.61$\pm$0.01\\
&$\sigma_{\rm{M_u}}$&1.07&1.05&1.18\\
&$\sigma_{\rm{\log(L[OII])}}$&0.38&0.45&0.46\\
&$r$&0.65&0.86&0.83\\
\midrule
\logaLOII\,=\,$A$\,\magg+$B$&$A$&-0.218$\pm$0.001&-0.342$\pm$0.001&-0.328$\pm$0.001&\\
&$B$&36.97$\pm$0.01&34.29$\pm$0.01&34.53$\pm$0.01\\
&$\sigma_{\rm{M_g}}$&1.11&1.08&1.15\\
&$\sigma_{\rm{\log(L[OII])}}$&0.38&0.45&0.46\\
&$r$&0.64&0.81&0.82\\
\midrule
\logaLOII\,=\,$A$\,log(age/yr)+$B$&$A$&---&---&-0.646$\pm$0.001&\\
&$B$&---&---&47.17$\pm$0.01\\
&$\sigma_{\rm{\log(age)}}$&---&---&0.54\\
&$\sigma_{\rm{\log(L[OII])}}$&---&---&0.46\\
&$r$&-0.44&-0.47&-0.76\\
\midrule
\logaLOII\,=\,$A$\,log(\Mstar/M$_{\odot}$)+$B$&$A$&---&0.563$\pm$0.001&---&\\
&$B$&---&35.70$\pm$0.01&---\\
&$\sigma_{\rm{log(M\star)}}$&---&0.52&---\\
&$\sigma_{\rm{log(L[OII])}}$&---&0.45&---\\
&$r$&0.54&0.64&0.03\\
\bottomrule
\end{tabular}\vspace{0.3cm}
\caption{Best-fit parameters of the linear scaling relations shown in Figs.\,\ref{fig:nz} and \ref{fig:propall}. All the \OII luminosities here are intrinsic and computed using the \GE code with input the instantaneous SFR for \SAGE and average SFR for \SAGE and \GAL.}
\label{tab:params}
\end{table*}


\subsection{L{\oiitit} versus broad-band magnitudes}
\label{sec:LO2_mags}
At a given redshift range, the broad-band magnitudes tracing the rest-frame UV emission of a galaxy are expected to be tightly correlated with the SFR and the production of emission line galaxies. The rest-frame UV slope ($1000 - 3000$\,\AA) at $z\sim 1$ is measured between the $u$ and the $g-$bands ($\sim 2000$\AA). As expected, these are the bands that correlate the most with both SFR and \OII luminosity for the sample under study.

The correlations between the broad-band $u$ and $g$ absolute magnitudes and the intrinsic \OII luminosity in \MD at $z\sim1$ are displayed in the first two columns of panels in Fig.\,\ref{fig:propall} together with DEEP2-FF observations. Data and all model galaxies show a good overlap in this parameter space. The observations populate a smaller region of the parameter space, while the SAMs extend down to lower SFR and \LOII values.
We over plot all the strong correlations (i.e. those with correlation coefficient $r\ge0.6$) as linear scaling laws with an associated scatter $\sigma_{\rm{y}}$. Their best-fit parameters ($A,\,B$) and correlation coefficients ($r$) can be found in Table\,\ref{tab:params}, where relations with $r<0.6$ have been omitted. We find both the $u$ and $g$ magnitudes to be tightly correlated with \LOII, and thus they have the potential to be used as proxies for the \OII luminosity, using the relations presented in Table \,\ref{tab:params}.

\subsection{L{\oiitit} versus age, metallicity and stellar mass}
\label{sec:LO2_age_Z}
We also study the dependence of the \OII luminosity on galaxy properties that are relevant to the \LOII and $(k+e)$ calculations: the age, metallicity, and stellar mass. 

The right column of panels in Fig.\,\ref{fig:propall} shows the relationship between the intrinsic \OII luminosity and the stellar mass in both DEEP2-FF and our model galaxies. In \SAGE we identify a correlation, but none is found for \SAG and \GAL model galaxies. The DEEP2-FF data do not exhibit any particular trend, maybe due to the narrow luminosity range that the sample covers.

In the third column of Fig.\,\ref{fig:propall}, we display the relationship between the intrinsic \LOII and age, which is mostly flat both in \MD and DEEP2-FF observations, with the latter showing a bimodal distribution. Only \GAL model galaxies exhibit an anti-correlation in the age-\LOII plane.

No correlation is found between the metallicity and \LOII for any of the models (this is not shown in Fig.\,\ref{fig:propall}). We conclude that none of the galaxy properties explored in this Section are good candidates as proxies for \LOII.


\subsection{From galaxy properties to L{\oiitit}}
\label{sec:test}
\begin{figure*}
\begin{center}
\includegraphics[width=\columnwidth]{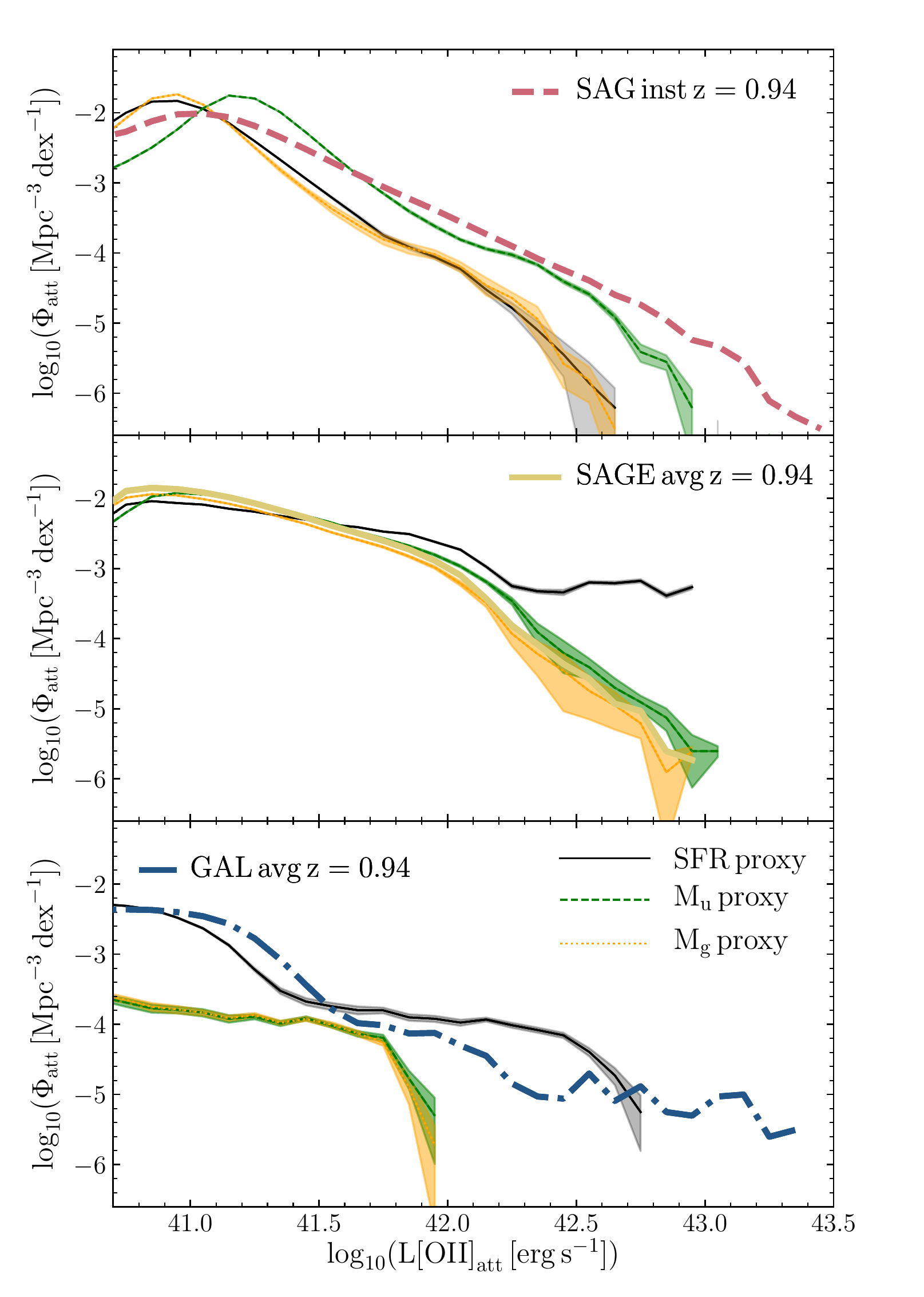}\quad
\includegraphics[width=\columnwidth]{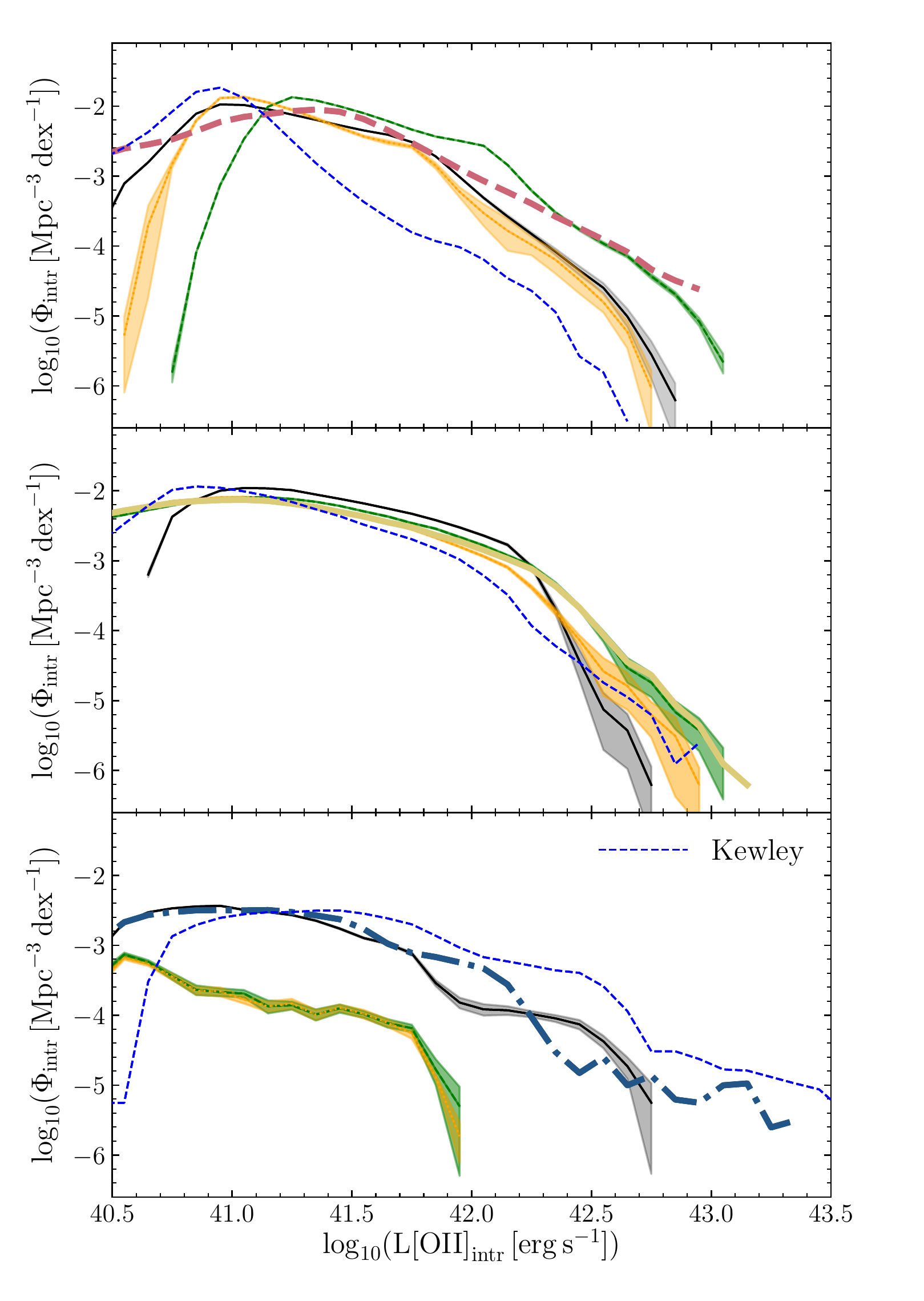}\vspace{-0.2cm}
\caption{\textit{Left column:} From top to bottom, attenuated \OII luminosity functions of the \SAG, \SAGE and \GAL model galaxies at $z\sim1$. We show as thick lines the results with \LOII computed using the \GE code described in Section\,\ref{sec:O2sams} with either instantaneous or average SFR and metallicity as inputs. We compare these results with the \LOII functions derived from the three \LOII proxies established above: SFR (solid, black line), M$_u$ (dashed, green) and M$_g$ (dotted, orange). The shaded regions represent the $\pm\sigma_y$ scatter in the proxy-\LOII linear scaling laws, which is  given in Table\,\ref{tab:params}. \textit{Right column:} Same as left column, but here the \LOII are intrinsic. The lines are colour-coded as the left panels. We show as blue dashed lines the results from the \citet{kewley04} conversion.} 
\label{fig:lumfuncproxies}
\end{center}
 \end{figure*}

The \LOII derived from the \GE code is tightly related to the SFR by construction, but we found it to be also tightly related with the broad-band $u$ and $g$ magnitudes ($r\ge0.64$, see Table~\ref{tab:params}). Here, we quantify the usability of the linear relations found as proxies to derive \LOII from average SFR and broad-band magnitudes. For this purpose, we compare the luminosity functions and galaxy clustering signal for \OII emitters selected using the aforementioned linear relations and the relations from Section\,\ref{sec:SFR-LO2}, with those obtained by coupling the SAMs with the \GE code (see Section\,\ref{sec:O2sams}).


\subsubsection{[O{\scshape{ii}}] luminosity functions}
\label{sec:LFev}
In the left column of Fig.\,\ref{fig:lumfuncproxies}, from top to bottom, we show the attenuated \OII luminosity functions of the \SAG, \SAGE and \GAL model galaxies at $z\sim1$. We compare the \LOII predictions from coupling the models with the \GE code (thick, coloured lines without error bars) with those from using the SFR (solid, black), \magu (dashed, green) and \magg (dot-dashed, orange) proxies established above and summarised in Table\,\ref{tab:params}. The shaded regions represent the effect of the scatter $\sigma_y$ on the proxy-\LOII relation and are derived from LFs estimated from 100 Gaussian realisations \textit{G($\sigma_y$,$\mu$)} with mean $\mu=$(SFR, \magu, \magg) and fixed scatter $\sigma_y=(\sigma_{\rm{SFR}}, \sigma_{\rm{M_u}}, \sigma_{\rm{M_g}})$ from Table\,\ref{tab:params}. 

The \OII luminosity functions derived from the proxies are strongly model dependent, with varying levels of success for each model and proxy, as can be seen in Fig.\,\ref{fig:lumfuncproxies}.
In \SAG, the \magu proxy produces a luminosity function which, in the \LOII range $10^{41.7}-10^{42.5}\,$erg\,s$^{-1}$,  is consistent with that derived from coupling the model with the \GE code, while the other two proxies are lower. In \SAGE, the \magg proxy returns a LF in very good agreement with the \GE estimate on all luminosity scales. \magu gives good agreement at \LOII$\lesssim10^{42}\,$erg\,s$^{-1}$, while beyond this value it slightly overestimates the number of \OII emitters. The SFR proxy is consistent with the \GE result at \LOII$\lesssim10^{41.7}\,$erg\,s$^{-1}$, while at higher \LOII values it overpredicts the luminosity function by $\sim1.5\,$dex.

The \LOII function based on the SFR proxy from \GAL is in reasonable agreement with that from coupling the model with \GE, while the magnitude proxies produce a lack of emitters on all luminosity scales ($\sim1.4\,$dex at $\sim10^{40.5}\,$erg\,s$^{-1}$, $\sim0.4\,$dex at $\sim10^{41.5}\,$erg\,s$^{-1}$ and $\sim1.8\,$dex at $\sim10^{42}\,$erg\,s$^{-1}$). Fig.\,\ref{fig:propall} shows that \GAL magnitudes are below those from DEEP2-FF. This discrepancy is likely to be the cause of the lack of \OII emitters.

In the right column of Fig.\,\ref{fig:lumfuncproxies}, we display the intrinsic \LOII functions colour-coded as the left panels. In \SAG and \SAGE model galaxies, the effect of dust attenuation is stronger at higher luminosities, while in \GAL it is more significant at \LOII$\lesssim10^{42}\,$erg\,s$^{-1}$. We overplot, as dashed, blue lines, the \OII luminosity functions obtained by applying the \citet{kewley04} conversion (Eq.\,\ref{eq:LC}) to each one of the model catalogues. 
This lies below (above) the other results in the bright end for \SAG and \SAGE (\GAL) model galaxies. The relation from \citet{kewley04} produces very different \LOII functions compared to the ones obtained from the SAM model galaxies coupled with the \GE prescription. This result highlights that the dispersion in the model gas metallicities is not the only source of the variation seen in the luminosity function in Fig.\,\ref{fig:lumfuncproxies}.

In this Section, we have investigated the impact in the \OII luminosity function of using the \LOII proxies established above. 
We find the \LOII proxies to be model-dependent and to overall result in either a lack or an excess of bright \OII emitters. 
These outcomes emphasise the inappropriateness of using simple relations to derive the \OII emission from global galaxy properties. In fact, besides introducing systematic uncertainties, they can also result in \OII luminosity functions with very different shapes depending which properties are used.

\subsubsection{Galaxy clustering}
\begin{figure}
\begin{center}
\includegraphics[width=\linewidth]{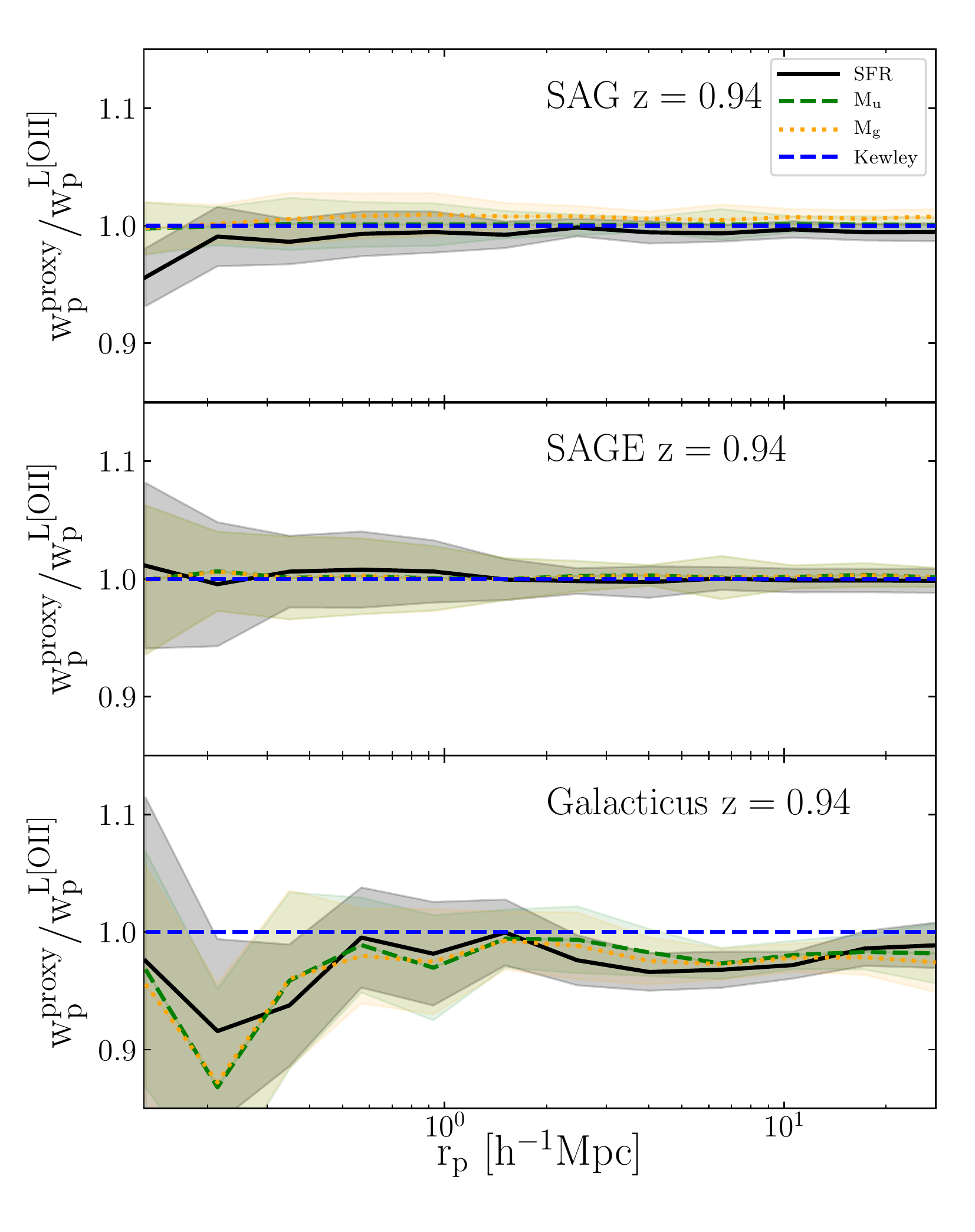}\vspace{-0.2cm}\vspace{-0.4cm}
\caption{Proxy-to-\LOII ratios of the projected two-point correlation functions of, from top to bottom, \SAG, \SAGE and \GAL model galaxies at $z\sim1$. The SAG \LOII is estimated using the \GE code with instantaneous SFR, while \SAGE and \GAL using the average quantity. Galaxies have been selected to have \LOII$>10^{40.4}\,$erg\,s$^{-1}$. The shaded regions represent the effect of the $\sigma_y$ scatter in the proxy-\LOII linear relations reported in Table\,\ref{tab:params}. These regions are the $1\sigma$ uncertainties derived from the co-variance of 100 Gaussian realisations with the \LOII proxy considered as mean and $\sigma_y$ as scatter. We over plot the \citet[]{kewley04} result as a dashed, blue line. }
\label{fig:CF}
\end{center}
\end{figure}

We further check how the clustering of our model ELGs is sensitive to an \OII luminosity selection, where \LOII is computed either from the \GE code, or the proxies established above. We consider \SAG, \SAGE and \GAL model galaxies at $z\sim1$ and impose on them a minimum luminosity threshold of \LOII$>10^{40.4}\,$erg\,s$^{-1}$. 

Fig.\,\ref{fig:CF} shows the ratios between the projected two-point correlation functions obtained from the proxy-to-\LOII relations and those derived from \LOII computed using the \GE code with instantaneous SFR (\SAG) or average SFR (\SAGE and \GAL). In Fig.\,\ref{fig:CF}, we also show the clustering of the data obtained using the conversion from \citet{kewley04} given in Eq.\,\ref{eq:LC}. For all the models, this clustering is in excellent agreement with the data derived from the \GE \LOII estimation. 

For the clustering we adopt the \citet[]{1993ApJ...412...64L} estimator and the two-point function code from \citet[]{Favole2015}. The shaded regions present the effect of the $\sigma_y$ scatter given in Table\,\ref{tab:params} in the proxy-\LOII linear relations. The dispersion is computed from the covariance of 100 Gaussian realisations with mean the desired proxy and scatter $\sigma_y$ (see Section\,\ref{sec:LFev} for further details). 

The clustering amplitude remains similar (within 12\%) for the different \LOII calculations in all the SAM considered. In particular, in \SAG and \SAGE galaxies, all the proxies agree within 5\% with the \GE and \citet{kewley04} results on all scales. On small scales, the SFR proxy in \SAG declines by 5\% and in \SAGE it shows some small fluctuations. In Galacticus, the clustering amplitude diminishes by up to $12\%$ ($4\%$) on small (intermediate) scales when assuming any proxy.

The two point correlation functions at $r_p>1h^{-1}$Mpc are consistent with each other, agreeing within the $1\sigma_y$ dispersion. 

We have investigated further the redshift evolution at $0.6<z<1.2$ and the dependence of different \LOII thresholds of the \MD clustering amplitude, both based on estimates from coupling the models with \GE and on the proxies above. 
In general, we find that increasing both the redshift and the \LOII thresholds, the galaxy number density decreases, resulting in a noiser clustering. Despite this increased noise, we find that model galaxies with \LOII$>10^{41}$erg s$^{-1}$ can be more clustered when \LOII is derived from proxies. We find variation among the different proxies used together with one of the three SAMs explored here. This possible dependency with \LOII should be taken into account when using proxies to create fast galaxy catalogues for a particular survey. 

Overall, we find that the \MD clustering signal is model-dependent. The linear bias is mostly unchanged, however differences are seen at small scales, below 1$h^{-1}$Mpc. The dispersion changes between the different proxies, with the SFR presenting the largest scatter, overall.

Our ELG clustering results show that simple \LOII estimates based on a linear relation with SFR are sufficient for modelling the large scale clustering of \OII emitters, even if they are not accurate enough to predict the \OII luminosity function.

\subsubsection{[O{\scshape{ii}}] ELG Halo Occupation Distribution}
\begin{figure}
\begin{center}
\includegraphics[width=\linewidth]{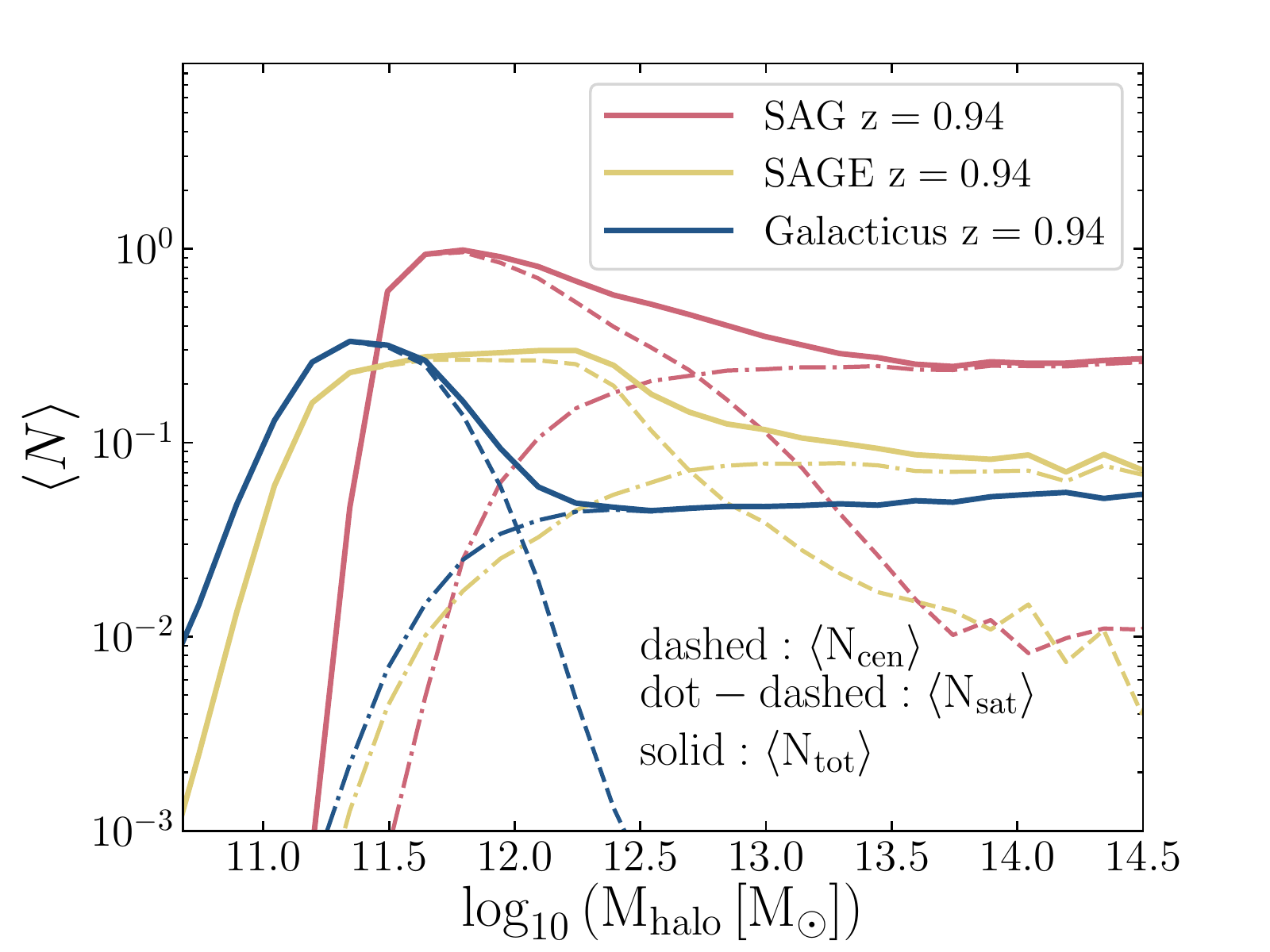}\vspace{-0.4cm}
\caption{Mean halo occupation distribution of the \SAG (salmon solid line), \SAGE (yellow solid line) and \GAL (blue solid line) model galaxies with \LOII$>10^{40.4}\,$erg\,s$^{-1}$ at $z\sim1$. The model \LOII has been computed using \GE with instantaneous SFR for \SAG galaxies and average SFR for \SAGE and \GAL. The contribution from central  galaxies is shown by dashed lines and that for satellites by dot-dashed lines.}
\label{fig:hod}
\end{center}
\end{figure}
\begin{figure}
\begin{center}
\includegraphics[width=\linewidth]{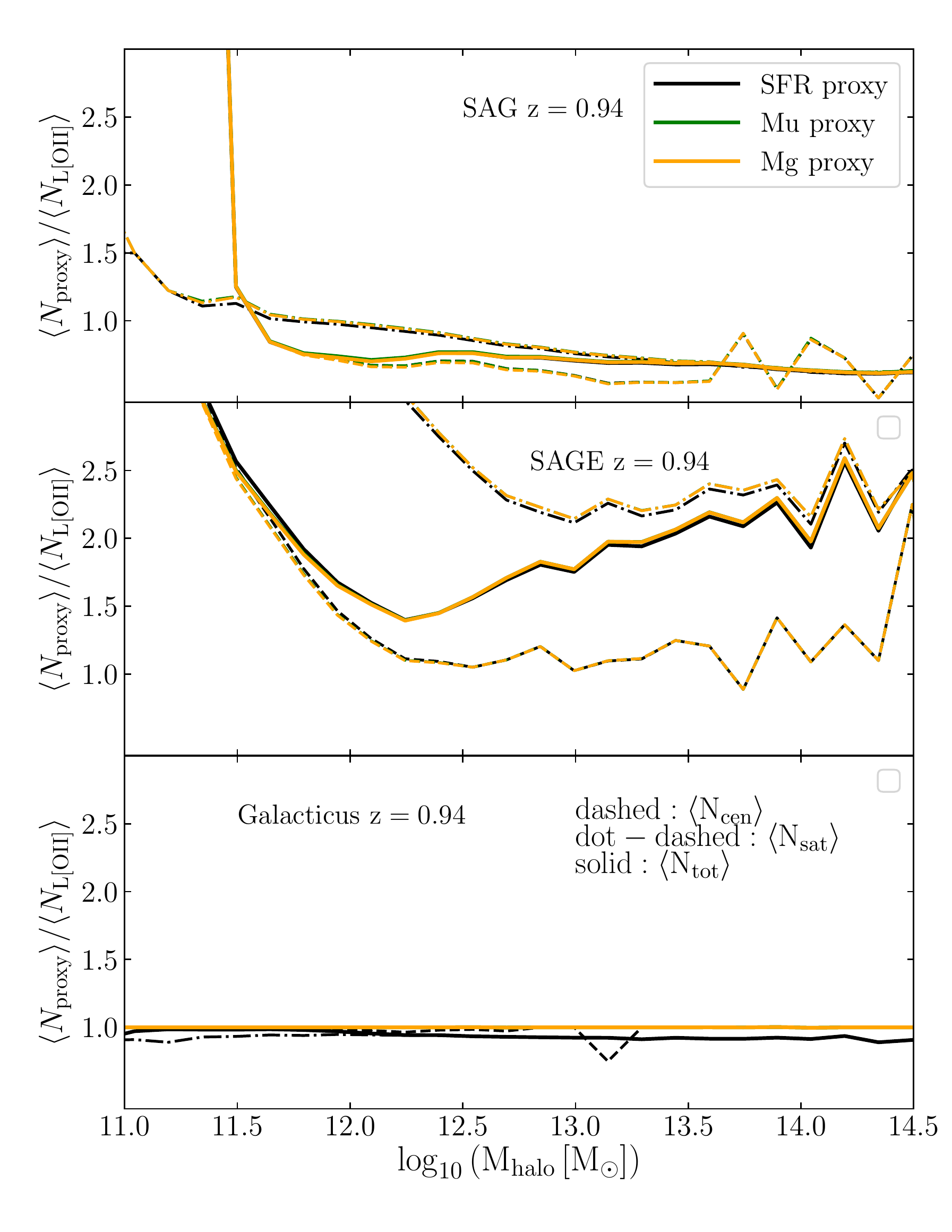}\vspace{-0.8cm}
\caption{Ratio between the HOD obtained from the \LOII calculated from the proxies indicated in the legend, and \LOII obtained using \GE. From top to bottom, results are shown for the \SAG, \SAGE and \GAL models.}
\label{fig:hodratios}
\end{center}
\end{figure}
In Fig.\,\ref{fig:hod}, we show the \MD mean halo occupation distribution (HOD) for model galaxies selected with  \LOII$>10^{40.4}\,$erg\,s$^{-1}$. Here, the model \OII luminosities have been calculated using the \GE code. We highlight contributions from central and satellite model galaxies. The shapes of the HODs are qualitatively consistent among the different models, with an asymmetric Gaussian for central galaxies, plus maybe a plateau, and a very shallow power law for satellite galaxies. A similar shape has been found using different models for either young or star-forming galaxies, selected in different ways~\citep{zheng2005,contreras13,cochrane2018eagle,violeta2018} and also in measurements derived from observations~\citep{geach12,cochrane2017,guo2018}. 

The shape of the HOD for central star-forming galaxies is very different from those selected with a cut in either rest-frame optical broad-band magnitudes or stellar mass, which is close to a smooth step function that reaches unity~\citep[e.g.][]{Berlind2002,Kravtsov2004}. As it can be seen in Fig.\,\ref{fig:hod}, the HOD of \MD central \OII emitters does not necessarily reach unity, i.e. it is not guaranteed to find an \OII emitter in every dark matter halo above a given mass. 

We find that the SAG HODs peak at higher halo masses compared to the other two SAMs. The mean halo masses predicted by the \SAG, \SAGE and \GAL model galaxies are in agreement with the results of \citet[]{favole16} for BOSS \OII ELGs at $z\sim0.8$ and \citet[]{2017MNRAS.472..550F} for SDSS \OII ELGs at $z\sim0.1$.

The HOD of \MD satellite \OII emitters is a very shallow power law, closer to a smooth step function. This is similar to what has been inferred for eBOSS \OII emitters~\citep{guo2018}, but very different to the findings using the {\sc galform} semi-analytical model~\citep{violeta2018}. This difference is most likely related to a different treatment of gas in this model, as the distribution of satellites in dark matter haloes of different masses is very sensitive to both the modelling of feedback and environmental processes.

In Fig.\,\ref{fig:hodratios}, we display the ratios between the \MD HODs selected in \LOII, where the luminosity is calculated from either using the \GE code or the proxies indicated in the legend. We find that the differences in the HODs from proxies and \GE are negligible for \GAL and less than $20\%$ for \SAG at \Mhalo$\gtrsim10^{12}{\rm M}_{\odot}$, while \SAGE shows differences above a factor 1.5 in most cases. The \LOII proxies behave very similarly, with negligible differences between them, except for the \GAL SFR proxy, which is slightly lower than the magnitude ones.

In summary, we find different levels of agreement with the \GE results depending on the model considered. However, the HOD remains almost unchanged when different \LOII proxies are assumed.


\section{Summary and conclusions}
\label{sec:discussion}
In this work, we have explored how the \OII luminosity can be estimated for semi-analytic models of galaxy formation and evolution using different methods: (i) by coupling the SAMs with the \GE code (Section\,\ref{sec:O2sams}) and (ii) using  simple relations between \LOII and global properties such as SFR, broad-band magnitudes and metallicity (Section\,\ref{sec:SFR-LO2}).

We have studied the following models from the \MD products \citep[][]{knebe2018}: \SAG \citep[][]{2018MNRAS.479....2C}, \SAGE \citep[][]{2016ApJS..222...22C} and \GAL \citep[][]{2012NewA...17..175B}. All these models are run on the MDPL2 cosmological simulation \citep[][]{klypin2016}. They were calibrated to a number of observations within $0<z<2$, and they produce SFR and stellar mass functions that evolve similarly to what is observed in this redshift range. 

Throughout this study, we have compared our model results with different observational data sets, including DEEP2-FF galaxies with absolute magnitudes (see Section~\ref{sec:deep2}). 

The \GE code to calculate nebular emission lines is publicly available and ideally uses instantaneous SFR as input. However, usually SAMs only output SFRs that are averaged over long time intervals, corresponding to the outputs of the underlying dark matter simulation. From the SAMs under study, only \SAG provides instantaneous SFRs. We have coupled the \GE code with the \SAG model using both instantaneous and average SFRs to study the impact that this choice has on the \LOII calculation in post-processing. Assuming as input for the \GE code either the instantaneous or the average SFR, we see a variation below 5\% for the dust attenuated \OII luminosity functions in the range $10^{41}-10^{42.2}\,$erg\,s$^{-1}$, and in the range $10^{41}-10^{3}\,$erg\,s$^{-1}$ for the intrinsic \OII luminosity functions. These ranges correspond to model ELGs with $1<$SFR (yr$^{-1}$M$_{\odot})<10^{1.5}$. At higher and lower SFRs, there is a larger discrepancy, $<50\%$, in the luminosity functions, when using either the average or the instantaneous SFR. Thus, we find that using average SFRs as inputs for \GE is a good approach when studying average galaxy populations.

The luminosity functions of the \MD with \LOII computed using the \GE algorithm are in good agreement with the DEEP2 and VVDS observations over the redshift range $0.6<z<1.2$. The \OII luminosity, SFR and stellar mass functions of {the} SAMs all consistently predict a smaller number of massive, star-forming emitters as the redshift increases.
The match we find in the \OII luminosity functions of model and DEEP2 galaxies, where the model \LOII values are computed using the \GE code, cannot guarantee that they are the same identical population of galaxies. In other words, we select the \SAG, \SAGE and \GAL model galaxies to best reproduce the characteristics of the observed DEEP2 \OII emitters. These selections return different levels of agreement in the explored parameter spaces, as shown in Figs.\,\ref{fig:proponly}, \ref{fig:propall}, \ref{fig:proponly_drawn}, \ref{fig:propdata}. A remarkable result from this study is that our model galaxies span the same regions as the observed ones, in all the parameter spaces under study, with overall consistent trends. This suggests that our modelling approach captures the  most important physical processes that shape the DEEP2 galaxy sample.

We have also investigated the viability of obtaining \LOII from simple relations with global galactic properties that are usually outputted by galaxy formation models. For this purpose, we use observationally derived relations~\citep[]{kewley04} and linear relations derived for each model. In particular, we explore the \LOII derived using the \GE code as a function of SFR, broad-band magnitudes, age and stellar mass. The SFR, both instantaneous and average, is the physical quantity that, by construction, is most correlated with the \OII luminosity (with correlation coefficients $r\geq0.80$ for all the SAMs). Such a tight correlation is well described by a linear scaling law with an associated scatter $\sigma_{\rm{log(SFR)}}$ that varies with \LOII (see Table\,\ref{tab:params}). Other valuable proxies to derive \LOII are the observed-frame $u$ and $g$ broad-band magnitudes, $M_{\rm u}$ and $M_{\rm g}$, which trace the rest-frame UV emission in our redshift range of interest. 

We test how feasible it is to use these correlations as proxies for \LOII by studying the evolution of the derived \OII luminosity functions, mean halo occupation distribution (HOD) and the galaxy clustering signal in \LOII thresholds.

The different methods explored to calculate \LOII result in a range of \OII luminosity functions. Taking into account the effect of the scatter in the SAG \LOII--proxy relations, the luminosity functions from the proxies (including the \citet[]{kewley04} relation from Eq.\,\ref{eq:LC}) are in reasonable agreement with the direct \GE estimates. The differences are larger for the relations derived from \magg and SFR in \SAG at all luminosities, for SFR in \SAGE at \LOII$>10^{42}\,$erg\,s$^{-1}$, and for the magnitude proxies in \GAL. At high luminosities, \LOII\, derived with most linear proxies result in a lack of bright emitters that increases with luminosity, but remains approximately constant with redshift. The \citet[]{kewley04} relation (Eq.\,\ref{eq:LC}) results in a lower number of bright \OII emitters compared to all the other methods to obtain \LOII in \SAG and \SAGE, and in a higher number in \GAL.

We find a large variation between the derived \OII luminosity functions among both {the SAMs} and the methods used to obtain \LOII. Thus, it is important to highlight that, despite the model SFR density evolution being in reasonable agreement with observations, simple relations based on global galaxy properties are not robust estimators for \LOII. 

We further test the use of simple relations to obtain \LOII for SAMs by measuring the galaxy two-point auto-correlation function for \OII emitters selected above a given \LOII threshold. We compare the clustering measured from the \OII proxies with direct predictions from the SAMs coupled with the \GE code and with the \citet{kewley04} relationship. The results vary from model to model and the largest fluctuations are seen below $1h^{-1}$Mpc. However, if we account for the effect of the scatter in the proxy-\LOII relation, the discrepancies reconcile with direct luminosity predictions. The large scale bias remains similar for all the models.

By increasing the \LOII threshold, the galaxy number density drops considerably resulting in a noisier clustering signal, which makes the comparison difficult. 
Despite this increased noise, we find that model galaxies with \LOII$>10^{41}$erg s$^{-1}$ can be more clustered when \LOII is derived from proxies (this depends both on the model and the proxy used). This possible dependency with \LOII should be taken into account when using proxies to create fast galaxy catalogues for a particular survey. 

There is no direct correspondence between a proxy resulting in a good luminosity function and providing a similar outcome for the clustering. 

We also test how the mean HOD of \OII emitters changes when assuming different proxies compared to the \GE code in the \LOII calculation of our SAMs. We find that the shape of the HOD is consistent with that expected for a star-forming population of galaxies. Quantitatively, the HOD is strongly model-dependent, and we find different levels of agreement between the proxies and the \GE results, in particular at \Mhalo$\gtrsim10^{12}{\rm M}_{\odot}$. However, the distributions remain substantially unchanged from one proxy to another for all the models under study.

Our results show that ELGs are different from SFR-selected samples and that the \LOII estimation needs more complex modelling than assuming a linear relation with SFR. Simple \LOII estimates are not accurate enough to predict direct statistics of \LOII, as the luminosity function, but they are sufficient for modelling the large scale clustering of \OII emitters. 

New-generation optical and infra-red surveys will provide enormous data sets with unprecedented spectroscopic precision and imaging quality. These observations, together with models of galaxy formation and evolution, will enable us to reach a complete and consistent understanding of both the Universe large scale structure, and the galaxy formation and evolution processes within dark matter haloes. 
In this context, simple derivations of \LOII might be adequate for the clustering above 1$h^{-1}$Mpc, although at least two simple approximations might be needed to determine the uncertainties.


\section{Data availability}
\label{sec:dataavail}
The data produced for this article are available at \url{http://popia.ft.uam.es/MultiDarkEmissionLines/
}. Here we provide the DEEP2--\textsc{Firefly} observations, both with and without dust attenuation, and the galaxy properties from the \SAG, \SAGE and \GAL models. For the latter, besides the \OII emission line, we also include the H$\alpha$, H$\beta$ and \OIII luminosities. 


\section*{Acknowledgments}
GF and VGP acknowledge support from the University of Portsmouth through the Dennis Sciama Fellowship award. VGP acknowledges support from the European Research Council under the European Union's Horizon 2020 research and innovation programme (grant agreement No 769130). DS is funded by the \textit{Spanish Ministry of Economy and Competitiveness} (MINECO) under the 2014 \textit{Severo Ochoa} Predoctoral Training Programme. DS also wants to thank the \textit{Mam\'ua Caf\'e Bar}-team for their kind (g)astronomical support. DS and FP acknowledge funding support from the MINECO grant AYA2014-60641-C2-1-P. AO acknowledges support from the Spanish Ministerio de Economia y Competitividad (MINECO) project No. AYA2015-66211-C2-P-2, and funding from the European Union Horizon 2020 research and innovation programme under grant agreement No. 734374. SAC acknowledges funding from {\it Consejo Nacional de Investigaciones Cient\'{\i}ficas y T\'ecnicas} (CONICET, PIP-0387), {\it Agencia Nacional de Promoci\'on Cient\'ifica y Tecnol\'ogica} (ANPCyT, PICT-2013-0317), and {\it Universidad Nacional de La Plata} (11-G124 and 11-G150), Argentina. CVM acknowledges CONICET, Argentina, for the supporting fellowship. AK is supported by the MINECO and the {\it Fondo Europeo de Desarrollo Regional} (FEDER, UE) in Spain through grant AYA2015-63810-P as well as by the MICIU/FEDER through grant number PGC2018-094975-C21. He further acknowledges support from the Spanish Red Consolider MultiDark FPA2017-90566-REDC and thanks Christopher Cross for sailing. ARHS acknowledges receipt of the Jim Buckee Fellowship at ICRAR-UWA. GF, VGP and DS wish to thank La Plata Astronomical Observatory for hosting the MultiDark Galaxies workshop in September 2016, during which this work was started. The authors thank the \textsc{Firefly} team and the anonymous referee for providing insightful comments. The analysis of DEEP2 data using the \textsc{firefly} code was done on the Sciama High Performance Compute cluster which is supported by the ICG, SEPNet and the University of Portsmouth (UK). The CosmoSim database used in this paper is a service by the Leibniz-Institute for Astrophysics Potsdam (AIP). The MultiDark database was developed in cooperation with the Spanish MultiDark Consolider Project CSD2009-00064. The authors gratefully acknowledge the Gauss Centre for Supercomputing e.V. (www.gauss-centre.eu) and the Partnership for Advanced Supercomputing in Europe (PRACE, www.prace-ri.eu) for funding the MultiDark simulation project by providing computing time on the GCS Supercomputer SuperMUC at Leibniz Supercomputing Centre (LRZ, www.lrz.de). The authors thank New Mexico State University (USA) and Instituto de Astrof\'isica de Andaluc\'ia CSIC (Spain) for hosting the \textsc{Skies \& Universes} database for cosmological simulation products. This work has benefited from the publicly available software tools and packages: \textsc{matplotlib}\footnote{\url{http://matplotlib.org/}\label{ftn:matplotlib}} 2012-2016 \citep[][]{Matplotlib}; \textsc{Python Software Foundation}\footnote{\url{http://www.python.org}\label{ftn:python}} 1990-2017, version 2.7., \textsc{Pythonbrew}\footnote{\url{https://github.com/utahta/pythonbrew}\label{ftn:pythonbrew}}; we use whenever possible in this work a colour-blind friendly colour palette\footnote{\url{https://personal.sron.nl/~pault/}\label{ftn:cb-friendly}} for our plots.

\bibliographystyle{mn2e}
\bibliography{./references}

\appendix

\section{SAG model galaxies selected from DEEP2-FF spline}
\label{sec:spline_results}
We study the properties of a subset of SAG model galaxies selected to reproduce the \LOII distribution of the DEEP2-\textsc{FF} data approximated by a spline fit in Fig.\,\ref{fig:spline}. We compare these model properties with the observational ones from DEEP2-\textsc{FF}.

\begin{figure}
\begin{center}\vspace{-0.4cm}
\includegraphics[width=1.05\linewidth]{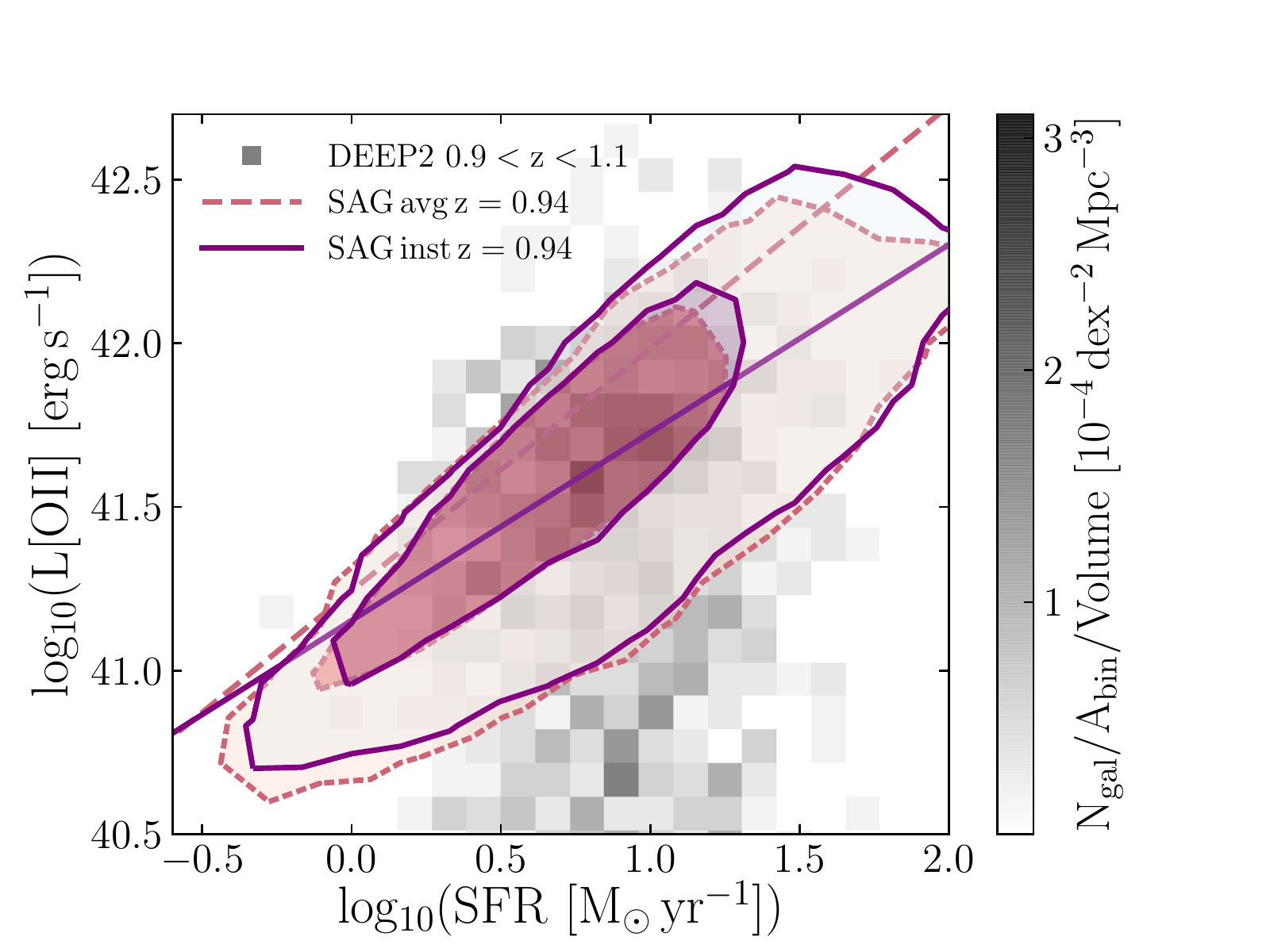}\vspace{-0.3cm}
\caption{Non-attenuated \OII luminosity as a function of SFR for the SAG model galaxies at $z\sim1$ (contours) and the DEEP2-FF observations at $0.9<z<1.1$ (grey, shaded squares). The colour bar represents the observed galaxy number density in each 2D bin normalised by the bin area in units of [dex$^{-2}$\,Mpc$^{-3}$]. Here the SAG model galaxies are selected following the spline fit to the observed DEEP2-FF \LOII distribution shown in Fig.\,\ref{fig:spline}. The model \LOII values are calculated by assuming instantaneous (solid, purple contours) and average (dashed, salmon) SFR as input for the \GE prescription. The innermost (outermost) model contours encompass 68\% (95\%) percent of the galaxy distributions. The diagonal lines represent the \LOII-SFR correlations, whose coefficients are reported in Tab.\,\ref{tab:params_prop_SAG_drawn}.}
\label{fig:SFRinstavg_drawn}
\end{center}
\end{figure}
\begin{table}\centering
\begin{tabular}{@{}lccccc@{}}\toprule
y=$A$\,x+$B$&  $A$&$B$&$\sigma_{\rm{y}}$&$r$ \\
\midrule
y=log$_{10}$(\LOII)&&&&\\
x=log$_{10}$(SFR$_{\rm{avg}}$)&0.741$\pm$0.002&41.24$\pm$0.01&0.41&0.92\\
x=log$_{10}$(SFR$_{\rm{inst}}$)&0.574$\pm$0.003&41.15$\pm$0.01&0.38&0.77\\
\midrule
y=\magu&&&&\\
x=log$_{10}$(SFR$_{\rm{avg}}$)&-2.021$\pm$0.006&-18.04$\pm$0.01&1.16&0.88\\
x=log$_{10}$(SFR$_{\rm{inst}}$)&-2.084$\pm$0.005&-17.88$\pm$0.01&1.17&0.90\\
\midrule
y=\magg&&&&\\
x=log$_{10}$(SFR$_{\rm{avg}}$)&-2.006$\pm$0.005&-18.96$\pm$0.01&1.11&0.92\\
x=log$_{10}$(SFR$_{\rm{inst}}$)&-2.032$\pm$0.005&-18.84$\pm$0.01&1.11&0.93\\
\midrule
y=log$_{10}$(M$_{\star}$)&&&&\\
x=log$_{10}$(SFR$_{\rm{avg}}$)&0.859$\pm$0.002&9.15$\pm$0.01&0.47&0.92\\
x=log$_{10}$(SFR$_{\rm{inst}}$)&0.846$\pm$0.002&9.11$\pm$0.01&0.47&0.90\\
\bottomrule
\end{tabular}
\caption{Best-fit parameters of the linear scaling relations found for \SAG model galaxies at $z=1$ and shown in Fig.\,\ref{fig:proponly_drawn}. The parameter $r$ is the correlation coefficient and $\sigma_y$ is the scatter in the $y$-axis. SAG galaxies  have been selected in \LOII randomly drawn from the DEEP2-FF spline fit shown in Fig.\ref{fig:spline}. }
\label{tab:params_prop_SAG_drawn}
\end{table} 
\begin{figure}
\begin{center}
\includegraphics[width=1.05\linewidth]{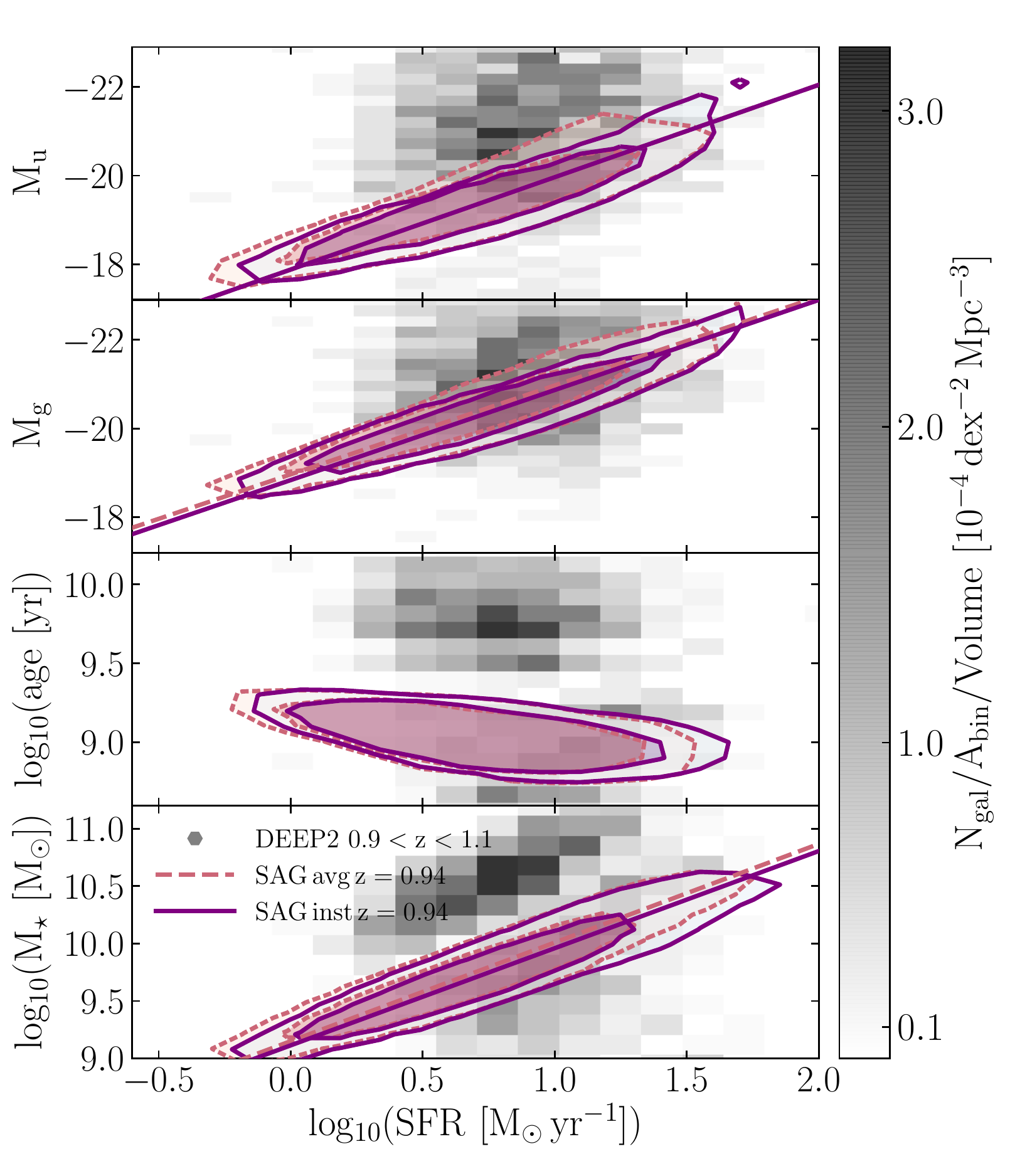}\vspace{-0.6cm}
\caption{From top to bottom: intrinsic magnitudes, ages and stellar masses as a function of star formation rate for \SAG model galaxies at $z\sim1$ (contours) and  DEEP2-FF observations at $0.9<z<1.1$ (grey, shaded squares). The model galaxies have been selected following the observed DEEP2-FF distribution approximated by a spline fit, as explained in Sec.\,\ref{sec:samplefinale}. The colour bar shows the number density of DEEP2 galaxies per bin area in units of [dex$^{-2}$\,Mpc$^{-3}$]. The dashed, salmon (solid, purple) contours represent the average (instantaneous) SFRs. The innermost (outermost) contours encompass 68\% (95\%) of the distributions. The diagonal lines are the linear fits showing the significant correlations, whose coefficients are reported in Table\,\ref{tab:params}, together with the best-fit parameters.}
\label{fig:proponly_drawn}
\end{center}
\end{figure}
Fig.\,\ref{fig:SFRinstavg_drawn} displays the SAG non-attenuated \OII luminosities computed from average and instantaneous SFRs as a function of SFR. The bimodality observed in Fig.\,\ref{fig:SFRinstavg}, where the model galaxies are selected by cutting at $\rm{SFR>0}$ and $\rm{log(M_{\star}/M_{\odot})>8.87}$, has now disappeared, but the discrepancy between the two sets of contours is larger, with the instantaneous correlation much shallower than the average one. At SFR$\gtrsim10^{1}$yr$^{-1}$M$_{\odot}$, the instantaneous SFR returns higher \LOII values compared to the average SFR. Instead, at SFR$\lesssim10^{0}$yr$^{-1}$M$_{\odot}$, the average contours reach fainter luminosities. Compared to the \SAG results based on simple SFR and stellar mass cuts (see Fig.\,\ref{fig:SFRinstavg}), here both sets of contours span a higher range of \LOII and SFR values. The \LOII-SFR correlation based on average (instantaneous) SFR is stronger (less strong) and with a steeper (shallower) slope compared to that for galaxies selected with simple cuts (compare Tabs.\,\ref{tab:params_prop_SAG} and \ref{tab:params_prop_SAG_drawn}), while the scatter is the same.

The DEEP2-FF observations in Fig.~\ref{fig:SFRinstavg_drawn} seem to span a narrower range in SFR {\bf and to go fainter in \LOII} compared to the model galaxy contours. However, we highlight that the low-luminosity observational tail has a very low-density of emitters ($\sim 10^{-4}$ in Fig.\,\ref{fig:SFRinstavg_drawn}).

In Fig.\,\ref{fig:proponly_drawn}, from top to bottom, we display the intrinsic $u$- and $g$-band absolute magnitudes, the age and stellar mass of the SAG model galaxies selected from the DEEP2-FF spline fit as a function of the average and instantaneous SFRs. Compared to the results based on simple cuts at SFR$\,>0$\,yr$^{-1}$\,M$_{\odot}$ and \Mstar$>10^{8.87}\,$M$_{\odot}$ (see Fig.\,\ref{fig:proponly}), here the correlations between SFR and magnitudes are steeper and M$_u$ shows a wider scatter in the $y$-axis. On the contrary, the correlation between SFR and stellar mass is shallower for \SAG galaxies drawn from the DEEP2-FF spline fit and with less scatter in the $y$-axis. The specific values of the correlation parameters and coefficients are reported in Table\,\ref{tab:params_prop_SAG_drawn}. Overall, \SAG galaxies selected from the spline fit reach brighter values of $u$- and $g$-band magnitudes compared to their counterparts based on simple SFR and stellar mass cuts (see Fig.\,\ref{fig:proponly}), which also extend down to fainter magnitudes and smaller stellar masses.
As already noticed in Fig.\,\ref{fig:proponly}, model galaxies have lower ages and stellar masses and they extend into larger SFR values, compared to the DEEP2-FF sample.

\section{Evolution of L{\oiitit} from instantaneous and average SFR}
\label{sec:appendix1}
We investigate further the redshift evolution of the small discrepancy generated in \LOII by assuming average instead of instantaneous SFR as input for the \GE code. In Section\,\ref{sec:instavg}, we have studied what happens at $z\sim1$, now we look over the redshift range $0.6<z<1.2$ to see if there is some evolution. 

\begin{figure}
\begin{center}\vspace{-0.2cm}
\includegraphics[width=\linewidth]{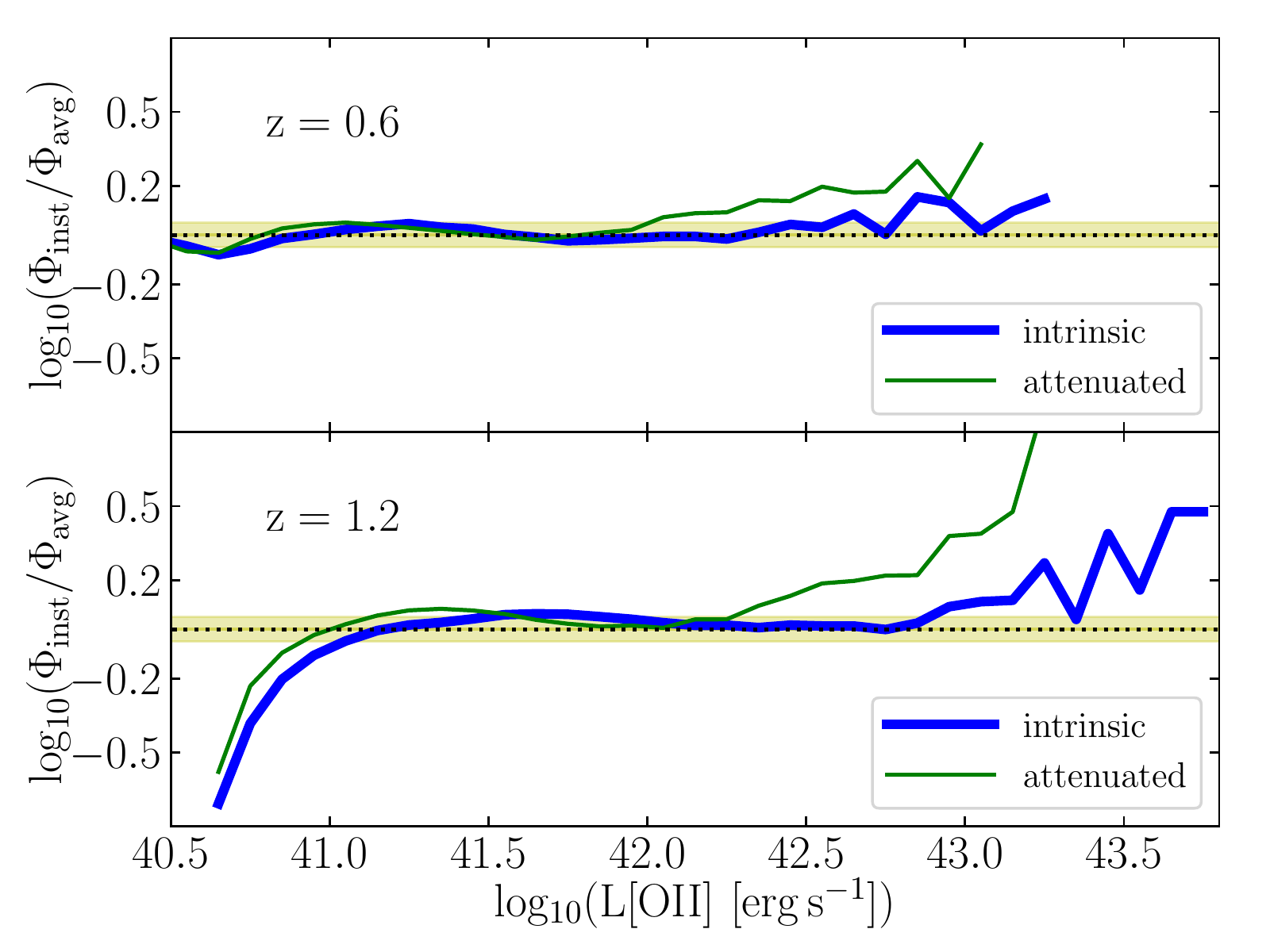}\vspace{-0.5cm}
\caption{Ratios between the SAG \OII luminosity functions (thick, blue lines: intrinsic LFs; thin, green: attenuated LFs) at different redshifts computed from average and instantaneous SFR using the method presented in\,\ref{sec:O2sams}. The yellow, shaded areas represent the 5\% confidence region.}
\label{fig:instavgevol}
\end{center}
\end{figure}

Fig.\,\ref{fig:instavgevol} compares the ratios of the intrinsic (thick, blue lines) and attenuated (thin, green) \OII luminosity functions obtained from average and instantaneous SFR at different redshifts. We have explored the entire range $0.6<z<1.2$ finding that, as the redshift increases, the instantaneous and average \LOII results tend to agree on a larger luminosity domain. Specifically, the 5\% agreement threshold (yellow, shaded region in the plot) is reached for the first time at $z=0.6,1.2$ by galaxies with attenuated \LOII$=10^{41.9},10^{42.3}\,$erg\,s$^{-1}$ and with intrinsic \LOII$=10^{42.6},10^{42.9}\,$erg\,s$^{-1}$, respectively. This result is independent on the presence of attenuation in the \OII luminosity. At $z\sim1.2$, we observe a larger discrepancy in both ratios in the faint region due to the larger effect of incompleteness.

\section{Global properties of \MD}
\label{sec:appendix2}
\begin{figure}
\begin{center}\vspace{-0.3cm}
\includegraphics[width=0.55\textwidth]{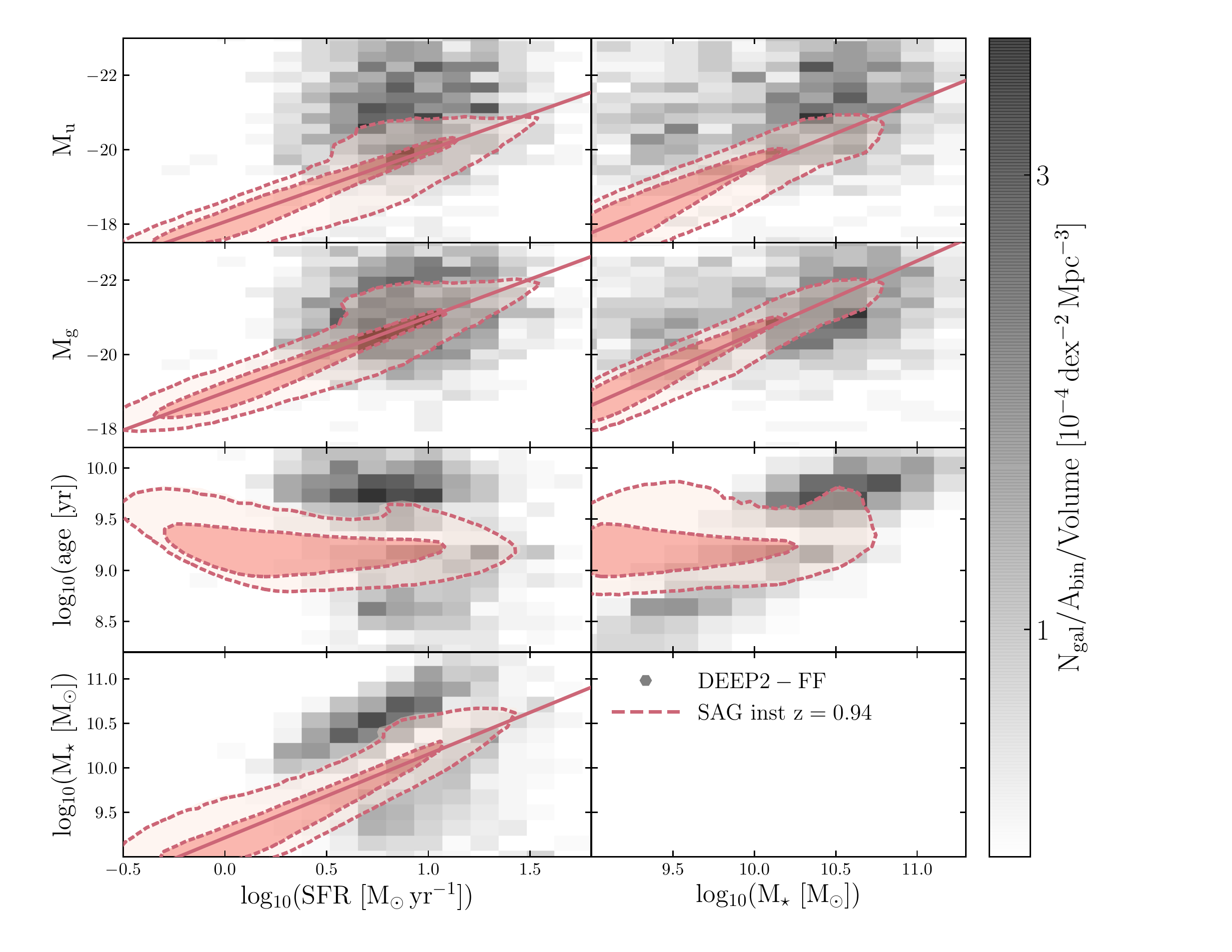}
\includegraphics[width=0.55\textwidth]{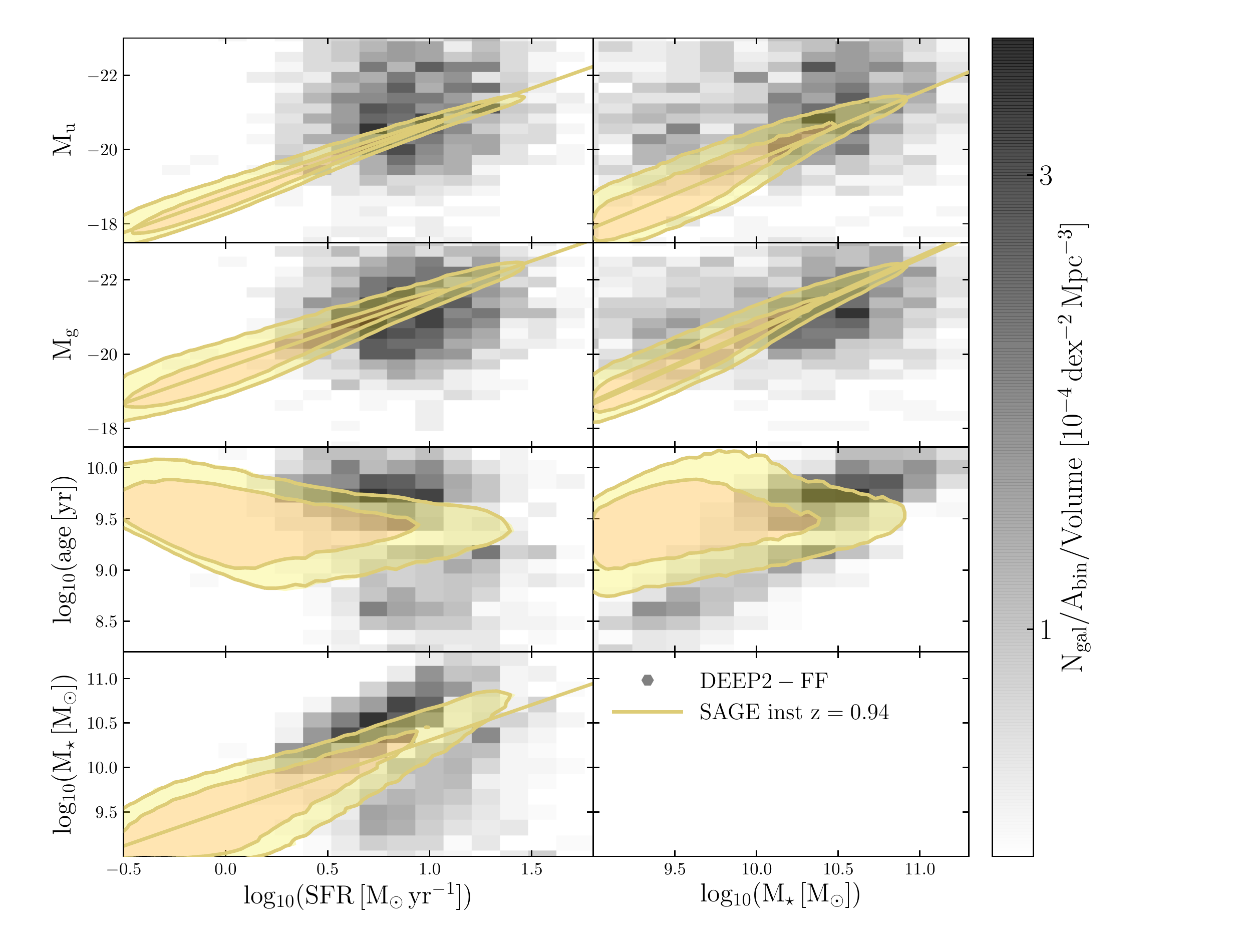}
\includegraphics[width=0.55\textwidth]{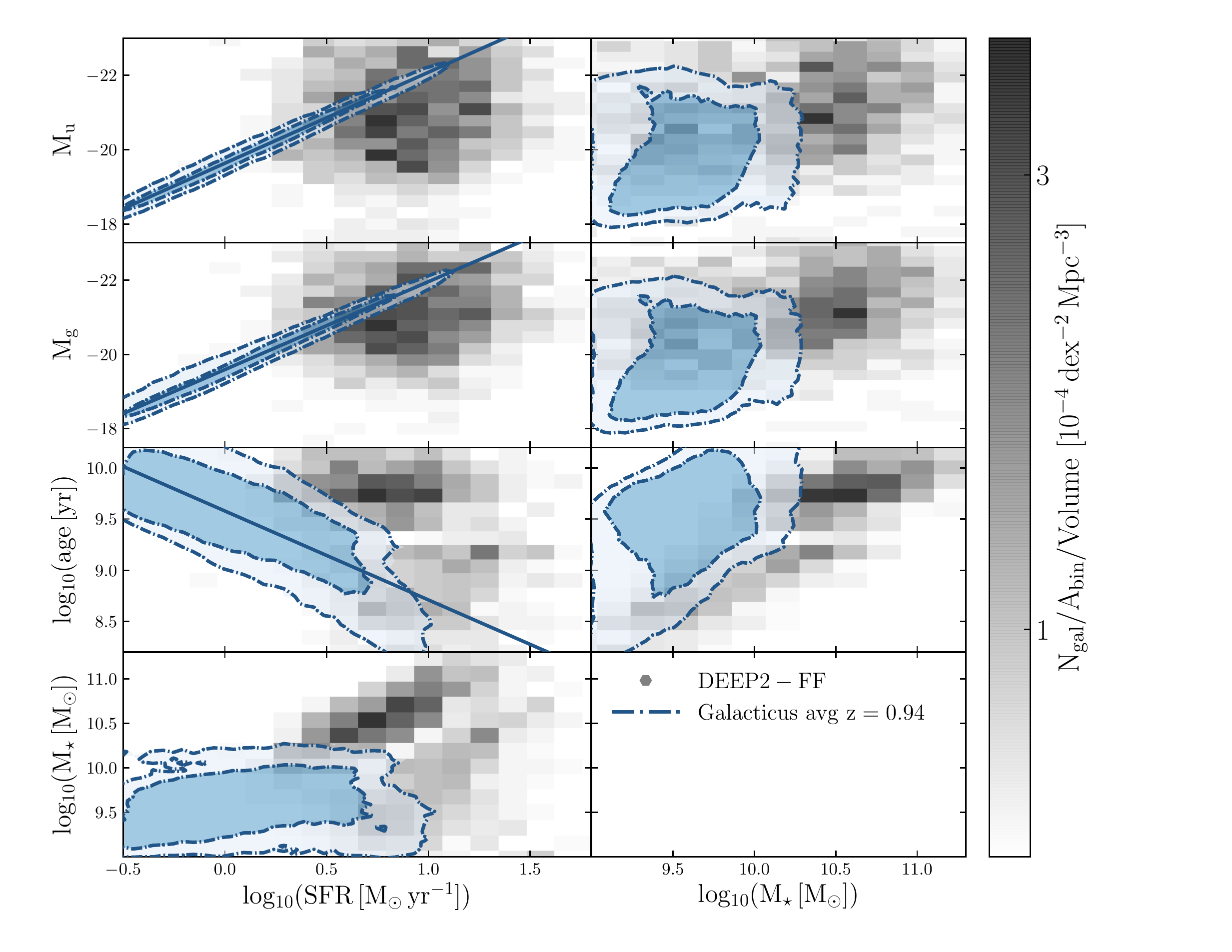}\vspace{-0.4cm}
\caption{Comparison of pairs of properties for \MD at $z=1$ (contours) and  DEEP2--FF observations at $0.9<z<1.1$ (grey, shaded squares). The colour bars show the number density of DEEP2 galaxies in each square. From top to bottom, we display \SAG, \SAGE and \GAL results. A minimum \OII flux cut of $\rm{5\times10^{-18}\,\,erg\,s^{-1}\,cm^{-2}}$ has been applied to both data and model galaxies. In each set of panels, from top to bottom, we compare broad-band $u$ and $g$ absolute magnitudes, age and stellar mass as a function of, from left to right, average SFR and stellar mass. The model contours, from inner to outer, represent 68\% and 95\% of the distributions. The diagonal lines are the linear fits showing the significant correlations, whose coefficients are given in Table\,\ref{tab:params}.}
\label{fig:propdata}
\end{center}
\end{figure}

\begin{table*}\centering
\begin{tabular}{@{}llcccc@{}}\toprule
\hspace{3cm}z=1&  & SAG& SAGE& Galacticus  \\
\midrule
\magu\,=$A$\,log$_{10}$(SFR/(M$_{\odot}$\,yr$^{-1}$))+$B$&$A$&-1.934$\pm$0.001&-1.941$\pm$0.001&-2.363 $\pm$0.001&\\
&$B$&-18.06$\pm$0.01&-18.74$\pm$0.01&-19.59$\pm$0.01\\
&$\sigma_{\rm{log(SFR)}}$&0.50&0.53&0.48\\
&$\sigma_{\rm{M_u}}$&1.07&1.05&1.15\\
&$r$&0.90&0.97&0.99\\
\midrule
\magg\,=$A$\,log$_{10}$(SFR/(M$_{\odot}$\,yr$^{-1}$))+$B$&$A$&-2.029$\pm$0.001&-1.916$\pm$0.001&-2.362$\pm$0.001&\\
&$B$&-18.98$\pm$0.01&-19.63$\pm$0.01&-19.59$\pm$0.01\\
&$\sigma_{\rm{log(SFR)}}$&0.50&0.53&0.48\\
&$\sigma_{\rm{M_g}}$&1.11&1.08&1.15\\
&$r$&0.91&0.94&0.99\\
\midrule
log$_{10}$(age/yr)\,=\,$A$\,log$_{10}$(SFR/(M$_{\odot}$\,yr$^{-1}$))+$B$&$A$&---&---&-0.869$\pm$0.002&\\
&$B$&---&---&9.58$\pm$0.1\\
&$\sigma_{\rm{log(SFR)}}$&---&---&0.48\\
&$\sigma_{\rm{age}}$&---&---&0.54\\
&$r$&-0.21&-0.34&-0.77\\
\midrule
log$_{10}$(M$_{\star}$/M$_{\odot}$)\,=$A$\,log$_{10}$(SFR/(M$_{\odot}$\,yr$^{-1}$))+$B$&$A$&0.939$\pm$0.001&0.794$\pm$0.001&---&\\
&$B$&9.21$\pm$0.01&9.51$\pm$0.01&---\\
&$\sigma_{\rm{log(SFR)}}$&0.54&0.52&---\\
&$\sigma_{\rm{log(M\star)}}$&0.50 &0.52&---\\
&$r$&0.87&0.81&0.18\\
\midrule
\magu\,=\,$A$\,log$_{10}$(M$_{\star}$/M$_{\odot}$)+$B$&$A$&-1.779$\pm$0.001&-1.820$\pm$0.002&---&\\
&$B$&-1.75$\pm$0.02&-1.52$\pm$0.01&---\\
&$\sigma_{\rm{log(M\star)}}$&0.54&0.52&---\\
&$\sigma_{\rm{M_u}}$&1.07&1.05&---\\
&$r$&0.89&0.89&0.15\\
\midrule
\magg\,=\,$A$\,log$_{10}$(M$_{\star}$/M$_{\odot}$)+$B$&$A$&-1.941$\pm$0.001&-1.951$\pm$0.001&---&\\
&$B$&-1.17$\pm$0.01&-1.15$\pm$0.02&---\\
&$\sigma_{\rm{log(M\star)}}$&0.54&1.08&---\\
&$\sigma_{\rm{M_g}}$&1.11&0.52&---\\
&$r$&0.94&0.94&0.18\\
\bottomrule
\end{tabular}\vspace{0.3cm}
\caption{Best-fit parameters of the linear scaling relations found in \MD at $z\sim1$ and shown in Fig.\,\ref{fig:propdata}. The parameter $r$ is the correlation coefficient and $\sigma_y$ is the scatter in the $y$-axis. The SFR values are instantaneous for \SAG and average for \SAGE and \GAL. We highlight that we do not quantify the correlation in the DEEP2-FF sample, since this calculation would require accounting for all the observational incompleteness effects, which goes beyond the aim of this work.}
\label{tab:params_prop}
\end{table*}

We compare pair properties in \MD and DEEP2-FF observations to better understand their mutual correlations. We then fit these dependences using linear scaling relations.
Fig.\,\ref{fig:propdata} displays, from top to bottom, the correlations between broad-band magnitudes, age and stellar mass as a function of SFR and stellar mass of the DEEP2-FF galaxies (grey, shaded squares, colour-coded according to their galaxy number density normalized by the 2D bin area) compared to the \MD (contours indicating the 68\% and 95\% of each distribution). Data and models overlap covering the brighter, more massive and more star-forming region of the parameter space. In particular, the \MD only cover the SFR range above the knee shown in Fig.\,\ref{fig:sfrf}. 

For such a small observational sample, it is difficult to establish and fit clear correlations among these quantities and between these quantities and \LOII (see also Fig.\,\ref{fig:propall}). In order to do this properly, one should account for all the DEEP2-FF incompleteness effects, which goes beyond the scope of our work. Here we show the comparison between the DEEP2-FF emitters and the \MD only to verify that our models cover the parameter space of the observational data set.

From the model point of view, we do find clear correlation among most of the physical quantities presented in Fig.\,\ref{fig:propdata}. Each set of panels shows the results for one model: from top to bottom we display \SAG, \SAGE and \GAL model galaxies. The relevant correlations ($r\ge0.6$) are represented as linear fits and the optimal parameters are reported in Table\,\ref{tab:params_prop}, together with their correlation coefficient ($r$) and the associated scatter in the $y$-axis ($\sigma_{\rm{y}}$). 

As expected, tight correlation is observed between the star formation rate and the broad-band $u$ and $g$ magnitudes that trace the rest-frame UV emission of a galaxy (see also Section\,\ref{sec:LO2_mags}). Tight correlation is observed also between the magnitudes and the stellar mass in all our model galaxies, except for \GAL. Overall, the DEEP2-FF observations and the \MD show a good overlap in the brighter, more star-forming and massive portion of any parameter space. All the model galaxies then extend further down in SFR, stellar mass and magnitudes.

Age does not correlate with SFR neither in the observations, nor in SAG and SAGE mocks. In \GAL, we observe an anti-correlation between age and SFR, meaning that older galaxies are more star-forming, as expected. Age does seem to correlate with stellar mass in DEEP2-FF, however this feature is not reproduced by any of our model galaxies. DEEP2-FF galaxies show a bimodal distribution in age and stellar mass, with an older, less star-forming, very massive population ($age\gtrsim10^{9.3}\,$yr; \Mstar$\gtrsim10^{10.3}\,$M$_{\odot}$) and a younger, more star-forming distribution with less massive galaxies. None of the model galaxies seem to reproduce this bimodality.

\SAG and \SAGE stellar masses are tightly correlated with their SFRs, but no dependence is observed in \GAL. While the DEEP2-FF quenched population is too sparse to identify any dependence in the stellar mass--SFR plane, the star-forming selection might show some correlation in the higher-mass end of the distribution. However, as already mentioned above, in order to correctly quantify this correlation, we should take into account the incompleteness effects in the data set, but this calculation goes beyond the aim of our analysis.
 We do not to show the dependence of the above quantities on metallicity since they do not correlate significantly in any of the model galaxies considered.

\end{document}